\documentclass[12pt]{article}

\pdfoutput=1 

\usepackage[T1]{fontenc} 

\usepackage{CJK}
\usepackage{mathtools}
\usepackage{tikz}
\usepackage{amssymb}
\usepackage{epsfig}
\usepackage{bbm}
\usepackage{bbold}
\usepackage{tabularx,booktabs}
\usepackage{graphicx}
\usepackage{hyperref}
\usepackage{romannum}
\usepackage{subfigure}
\usepackage{graphics,graphpap}
\usepackage{graphicx}
\usepackage{amsmath,exscale,amsthm,bm}
\usepackage{graphicx}
\usepackage{yfonts}
\usepackage{mathrsfs}
\usepackage[toc]{appendix}
\usetikzlibrary{trees}
\usetikzlibrary{positioning,arrows}
\usetikzlibrary{decorations.pathmorphing}
\usetikzlibrary{decorations.markings}
\usetikzlibrary{decorations.text}
\usetikzlibrary{fit,chains,calc,shapes.geometric,intersections}
\usepackage{natbib}
\setcitestyle{square,comma,numbers,sort&compress}
\usepackage{circledsteps}

\begin{document}

\def\mc#1{\mathcal#1}
\def\a{\alpha}
\def\b{\beta}
\def\c{\chi}
\def\d{\delta}
\def\e{\epsilon}
\def\f{\phi}
\def\g{\gamma}
\def\h{\eta}
\def\i{\iota}
\def\j{\psi}
\def\k{\kappa}
\def\la{\lambda}
\def\m{\mu}
\def\n{\nu}
\def\o{\omega}
\def\p{\pi}
\def\q{\theta}
\def\r{\rho}
\def\s{\sigma}
\def\t{\tau}
\def\u{\upsilon}
\def\x{\xi}
\def\z{\zeta}
\def\D{\Delta}
\def\F{\Phi}
\def\G{\Gamma}
\def\J{\Psi}
\def\L{\Lambda}
\def\O{\Omega}
\def\P{\Pi}
\def\Q{\Theta}
\def\S{\Sigma}
\def\U{\Upsilon}
\def\X{\Xi}

\def\ve{\varepsilon}
\def\vf{\varphi}
\def\vr{\varrho}
\def\vs{\varsigma}
\def\vq{\vartheta}

\newcommand{\vev}[1]{\langle #1 \rangle}
\def\dg{\dagger}                                     
\def\ddg{\ddagger}                                   
\def\wt#1{\widetilde{#1}}                    
\def\mt{\widetilde{m}_1}
\def\mti{\widetilde{m}_i}
\def\mtj{\widetilde{m}_j}
\def\rt{\widetilde{r}_1}
\def\mtt{\widetilde{m}_2}
\def\mttt{\widetilde{m}_3}
\def\rtt{\widetilde{r}_2}
\def\mb{\overline{m}}
\def\VEV#1{\left\langle #1\right\rangle}        
\def\be{\begin{equation}}
\def\ee{\end{equation}}
\def\ds{\displaystyle}
\def\ra{\rightarrow}

\def\bea{\begin{eqnarray}}
\def\eea{\end{eqnarray}}
\def\NO{\nonumber}
\def\Bar#1{\overline{#1}}
\def\ylz{\textcolor{red}}

\title{
\vspace*{1mm}
{\bf Extended $SO(10)$-inspired leptogenesis}

\author{
{\Large Pasquale Di Bari and  Xubin Hu}
\\
{\it Physics and Astronomy}, 
{\it University of Southampton,} \\
{\it  Southampton, SO17 1BJ, U.K.} 
\\
}
}
\maketitle \thispagestyle{empty}
\pagenumbering{arabic}

\begin{abstract}
We consider seesaw models with hierarchical Dirac neutrino masses. 
We show that, barring well known 
fine-tuned crossing level solutions and given current low energy neutrino experimental data, 
these always lead to a very hierarchical heavy neutrino spectrum. 
If the Dirac neutrino masses are assumed not to significantly differ 
from the up-quark masses, then the intermediate seesaw neutrino ($N_2$) mass falls in the correct 
range for its decays to  explain the matter-antimatter asymmetry of the universe via ($N_2$-)leptogenesis,  
while the lightest seesaw neutrino decays give a negligible contribution. 
We show that these conclusions are valid independently of any additional condition on the Dirac neutrino mass matrix. 
In this way we extend the usual $SO(10)$-inspired leptogenesis scenario, assuming the mismatch between neutrino 
and charged lepton Dirac mass matrices at the level of the CKM matrix. 
We find that solutions for inverted ordered neutrino masses become viable.  We still find a lower bound on the lightest neutrino mass, 
$m_1 \gtrsim 10^{-5}\,{\rm eV}$, but this is relaxed by two orders of magnitude.   We also consider the special subset 
of solutions realising strong thermal leptogenesis within this extended scenario, finding that inverted ordering remains incompatible.
Our results show how the emergence of the $N_2$-leptogenesis scenario within seesaw models is much more general than thought so far.  
\end{abstract}

\newpage

\section{Introduction}

After almost thirty years from their discovery \cite{Super-Kamiokande:1998kpq}, neutrino oscillations still provide the only  
evidence of physics beyond the standard model (SM) in particle physics experiments. They unambiguously establish neutrino 
masses, a missing ingredient in the SM. Moreover, thanks to a relentless experimental advance,
they still represent, together with absolute neutrino mass scale experiments, a leading phenomenology to understand how to extend 
the SM in order to explain the origin of neutrino masses.  The minimal (and therefore most attractive) way is via
type-I seesaw mechanism \cite{seesaw}, where right-handed (RH) neutrinos with Yukawa couplings to lepton and Higgs doublets and a 
Majorana mass term are added to the SM Lagrangian.
Due to the many additional parameters, low energy neutrino experiments alone do not 
overconstrain the seesaw parameter space, in fact providing information only on half of the parameters 
in case of seesaw models with three right-handed neutrinos. One has then either to rely on additional theoretical information reducing 
the number of  independent parameters in order to increase the predictive power and/or
search for additional observables, depending on seesaw parameters, that might be measured experimentally.  
In this respect, an important complementary phenomenological tool 
that provides an additional measurable observable is represented by cosmological observations.
Specifically, these are telling us that our observable universe is matter-antimatter asymmetric. 
The matter-antimatter asymmetry of the universe itself then provides the additional measured observable that
 needs to be explained in addition to low energy neutrino parameters.  This can be done with leptogenesis \cite{fy}, 
 a baryogenesis model where the asymmetry is in first place generated in the form of lepton number
 by the decays of the new heavy neutrino states predicted by the seesaw mechanism (seesaw neutrinos). 
 If the decays occur in the early universe at sufficiently high temperatures, sphalerons reprocess part of the $B-L$ asymmetry into
 a baryon asymmetry that has to reproduce the observed one \cite{manton,kuzmin,klebnikov,Harvey:1990qw}. 

In low-scale seesaw models, the seesaw neutrino masses are light enough that can be tested, by definition, in laboratory experiments in different ways \cite{Deppisch:2015qwa}. 
For example, they can be tested either at the LHC with displaced vertex searches if they are lighter than gauge bosons \cite{Izaguirre:2015pga}, 
and/or, if lighter than bottom quark mass, in beam dump experiments such as SHiP \cite{SHiP:2018xqw}.  The lack of evidence of new physics in collider experiments 
has so far  yielded constraints \cite{Drewes:2019fou}. These also start to marginally constraint allowed regions in low-scale leptogenesis scenarios. 
Future HL-LHC and SHiP data will provide more significant tests and be able to rule out large portion of the allowed region in the space of parameters.
 
In the case of more traditional high scale seesaw models, the explanation of the matter-antimatter asymmetry with leptogenesis is still not a sufficient condition to produce definite predictions in combination with low energy neutrino data. One can also try to explain simultaneously dark matter, obtaining a unified model of neutrino masses and origin of matter in the universe. 
In this case, one could expect to observe some signal either in very high energy neutrinos or in gamma rays \cite{Anisimov:2008gg,DiBari:2016guw}.
The recent discovery of gravitational waves \cite{LIGOScientific:2016aoc} has also opened new opportunities 
to test high scale leptogenesis models related to the possibility of detecting a
primordial cosmological background of gravitational waves
\cite{Buchmuller:2019gfy,Dror:2019syi,DiBari:2020bvn,DiBari:2021dri,Fu:2022lrn}.

While waiting to discover such possible new observables, another quite effective strategy can be pursued.
Seesaw models can traditionally and naturally be embedded within well motivated grand-unified models. 
In particular, $SO(10)$ models \cite{SO10} represent an elegant way to realise the seesaw mechanism, 
since they predict three RH neutrinos, one per each family. They also produce relations among fermion mass matrices, including the RH neutrino Majorana mass matrix, 
that reduce the number of independent parameters. Fitting all measured SM and neutrino parameters, searching for a realistic model,
is a challenging task that can lead to interesting nontrivial predictions on the unknown neutrino parameters. 
If the matter-antimatter asymmetry calculated with leptogenesis is also included, 
finding such fits becomes even more challenging and predictive. 

On the other hand, there are many different $SO(10)$ models. They can be even obtained from bigger groups, such as $E_6$ \cite{Babu:2024ecl},
or one can also combine grandunified symmetries with discrete flavour groups \cite{Morisi:2007ft,Hagedorn:2010th,DiBari:2015oca}. 
For each model one can get a different set of predicted values for the unknown neutrino parameters. 
In an attempt to understand whether there can be general common features in such predictions, that can be regarded as a distinctive imprint of these models, 
the class of so-called $SO(10)$-inspired models \cite{SO10inspired} are defined by a minimal condition that all these models have in common: 
the neutrino Dirac mass matrix is assumed to be not too dissimilar from the up-quark mass matrix, a typical outcome of minimal $SO(10)$ models.  
This condition can be regarded as the combination of two different conditions:
\begin{itemize}
\item[(I)] The neutrino Dirac masses, the eigenvalues of the neutrino Dirac mass matrix, are order-of-magnitude wise equal to the up quark masses.
\item[(II)] The unitary matrix, that we denote by $V_L$, that transforms the left-handed (LH) 
neutrinos from the flavour basis, where the charged lepton mass matrix is diagonal, 
to the (neutrino) Yukawa basis, where the neutrino Dirac mass matrix is diagonal, does not significantly differ from the CKM matrix in the quark sector. 
\end{itemize}
From  condition (I) one can derive analytical expressions for the RH neutrino masses in terms of the low energy neutrino parameters, the Dirac neutrino masses and the unitary matrix $V_L$. 
Imposing condition (II), it has been shown that, plugging current information on low energy neutrino parameters, the RH neutrino mass spectrum is very hierarchical 
and in particular the lightest RH neutrino mass $M_1 \ll 10^{9}\,{\rm GeV}$.
From the well known lower bound on the heavy neutrino mass obtained imposing successful leptogenesis \cite{di,cmb}, this implies  that the lightest RH
neutrino decays cannot explain the observed asymmetry. On the other hand, the second lightest RH neutrino has a  mass $M_2 \sim 10^{11}\,{\rm GeV}$,
just of the right order to yield successful $N_2$-leptogenesis, if the interference with the heaviest seesaw neutrino ($N_3)$ is sufficiently strong. 
This implies that the $N_3$ neutrino mass $M_3$ cannot be too much larger than $\sim 10^{16}\,{\rm GeV}$ and, from the seesaw formula, this yields a lower bound on the lightest neutrino
mass $m_1 \gtrsim 1\,{\rm meV}$ \cite{riotto1}. This is the $SO(10)$-inspired leptogenesis scenario, a direct consequence of 
$SO(10)$-inspired conditions, that can successfully reproduce the observed asymmetry. 

The $N_1$-washout of the asymmetry from $N_2$-decays
is strongly reduced when flavour effects are taken into account  \cite{vives}.
Therefore, these play a crucial role in determining the existence of solutions for successful $SO(10)$-inspired leptogenesis \cite{riotto1}. 
These solutions correspond to 
 holes in the lightest RH neutrino washout opened by flavour effects. They imply meaningful constraints 
 on the low energy neutrino parameters \cite{riotto2}.  In particular, with the current upper bound on the atmospheric neutrino mixing angle 
 and on the absolute neutrino mass scale from cosmological observations,  there are no solutions for inverted ordered (IO) neutrino masses \cite{full}. This implies that  $SO(10)$-inspired leptogenesis strictly predicts normal ordering (NO). 
 This is interesting, in light of the expected results from the  JUNO experiment that, also in combination with the other neutrino oscillation parameters, should determine in a few years the neutrino mass ordering. Preliminary results from JUNO combined with the other neutrino oscillation experiments already tend to favour NO at $2.3\sigma$ \cite{talks_nu26+pas26}, in agreement 
 with atmospheric neutrino data favouring NO at $92.3\%$ C.L. ($\sim 1.8\s$) \cite{Super-Kamiokande:2023ahc}. 
 This is certainly  encouraging but of course we need to wait for more data to draw a firm conclusion on neutrino mass ordering. 
 Another very interesting constraint, is that in the range $10\,{\rm meV} \lesssim m_1 \lesssim 30\,{\rm meV}$,
 a range that cosmological observations are currently start to test, 
 there is an upper bound on the atmospheric neutrino mixing angle implying the first octant. 

An intriguing feature of $SO(10)$-inspired leptogenesis is that a subset of the solutions realises so-called {\em strong thermal leptogenesis}. 
In this case the final asymmetry is independent of the initial conditions. In particular, an initial large pre-existing $B-L$ asymmetry, as large as $N_{B-L} \sim 0.1$,  is washed-out, while part of the asymmetry generated by $N_2$-decays can survive (in the tauon flavour) \cite{problem}. This scenario requires quite 
special conditions to be realised and it is highly non-trivial that such a subset of solutions exists within $SO(10)$-inspired leptogenesis \cite{STSO10}. 
In this case, all constraints on low energy neutrino parameters become much more stringent and the scenario highly predictive. 
The allowed range on the absolute neutrino mass scale is  $10\,{\rm meV} \lesssim m_1\lesssim 30\,{\rm meV}$. This translates into a similar window
on the effective neutrinoless double beta decay neutrino mass, testable in next generation experiments, such as KamLAND2-ZEN \cite{Nakamura:2020szx}, LEGEND1000 \cite{LEGEND:2021bnm}, CUPID \cite{CUPID:2025avs}.
In this range, as mentioned, the atmospheric neutrino mixing angle has to lie in the first octant. In addition, the strong thermal
$SO(10)$-inspired leptogenesis solution requires the $C\!P$ violating Dirac phase in neutrino oscillations to lie approximately in the fourth quadrant. 
This set of predictions is compatible with current data.  It is also remarkable that the strong thermal $SO(10)$-inspired solution has correctly predicted a nonzero reactor mixing angle $\theta_{13}$ \cite{DESY}.

This favourable phenomenological picture encourages further investigations on how robust these constraints are. First of all with respect to theoretical uncertainties, but  also to the same exact formulation of the $SO(10)$-inspired conditions. A systematic study of the theoretical uncertainties has been started accounting for flavour coupling from spectator processes \cite{DiBari:2025zlv}. 
This investigation has shown that new muonic (and some very marginal electronic) solutions appear but that overall the picture remains unchanged. 
The subset of solutions realising the strong thermal condition survives, in fact the allowed region even enlarges. 

In this paper we extend the $SO(10)$-inspired leptogenesis relaxing $SO(10)$-inspired condition (II) and deriving the full set of solutions satisfying successful
leptogenesis for an arbitrary matrix $V_L$. Some partial results in this direction were already conducted in \cite{STSO10}
and \cite{DiBari:2020plh}, but the angles in the matrix $V_L$ were still limited (all three) to be not larger than the Cabibbo angle in the CKM matrix. It was found
in \cite{DiBari:2020plh} that for the largest allowed  values of the angle $\theta_{23}^L$ in $V_L$, the lower bound on $m_1$ was getting relaxed to $\sim 0.2\,{\rm meV}$. 
Interestingly, in the search of a realistic fit within a minimal $SO(10)$-model \cite{Babu:2024ahk}, a best fit was found for $m_1 \simeq 0.04\,{\rm meV}$. This is not in contradiction with the result found in \cite{DiBari:2020plh} since for this best fit one has a maximal value $\theta_{23}^L \simeq 45^\circ$. 
Moreover, a very marginally statistically allowed fit for IO was also found. On the other hand, the other two angles $\theta_{12}^L$ and $\theta_{13}^L$ are smaller than the respective angles in the CKM, thus respecting $SO(10)$-inspired condition (II). 
Importantly, the RH neutrino mass spectrum is still of the type leading to $N_2$-leptogenesis for the explanation of the observed asymmetry.

These results prompt our more general investigation on how the solutions extend when $V_L$ is arbitrary. Very importantly,
we find that the RH neutrino mass spectrum still remains very hierarchical, of the same kind 
that is found imposing $I \lesssim V_L \lesssim V_{\rm CKM}$, thus necessarily requiring $N_2$-leptogenesis for the explanation of the asymmetry. 
We still find that crossing level solutions imply a very high level of fine-tuning in the seesaw formula and we are able to give a simple analytical explanation. 
The $N_2$-leptogenesis scenario is then an unavoidable consequence if one wants to reproduce the observed asymmetry with leptogenesis,
assuming Dirac neutrino masses not too dissimilar (order-of-magnitude wise) from up quark masses and barring high fine-tuned solutions. 
The only way to change this conclusion is either to have a very
low light Dirac neutrino mass $m_{D1} \ll m_{\rm up} \sim 1\,{\rm MeV}$ \cite{DiBari:2015svd}, such that $M_1$ becomes lower than the sphaleron out-of-equilibrium temperature $T_{\rm sph}^{\rm out} \simeq 132\,{\rm GeV}$ \cite{DOnofrio:2014rug}, and in this case one would fall into 
an effective two RH neutrino scenario \cite{Antusch:2011nz}, or a very  heavy one, $m_{D1} \gg m_{\rm up} \sim 1\,{\rm MeV}$,
such that $M_1 \gtrsim 10^{9}\,{\rm GeV}$, recovering the traditional $N_1$ leptogenesis scenario.

Another important result of our investigation is that within such a general extension of $SO(10)$-inspired leptogenesis, 
a lower bound $m_1 \gtrsim 0.1 \, {\rm meV}$ still exists, though relaxed by about two orders of magnitude. 
However, as we will point out, the bulk of solutions is still comprised in a range $1\,{\rm meV}  \lesssim m_1 \lesssim 30\,{\rm meV}$ 
and analogously for $m_{ee}$.

The paper is structured in the following way. In Section 2 we review $SO(10)$-inspired conditions and $SO(10)$-inspired leptogenesis, 
showing the main results updated using latest experimental data. 
In Section 3 we introduce extended $SO(10)$-inspired models and leptogenesis.  
We show the numerical results of scans in different scatter plots and for different range of values of the three angles in $V_L$.
We show how constraints obtained in the traditional scenario get relaxed and even IO solutions become viable.
We also provide some analytical insight on the numerical results. Finally, in Section 4 we draw our conclusions.

\section{$SO(10)$-inspired leptogenesis}

In this section we shortly review how from $SO(10)$-inspired conditions one necessarily requires,
barring fine-tuned crossing level solutions, the $N_2$-leptogenesis scenario for the explanation of the 
matter-antimatter asymmetry of the universe.
We also update the results using the latest low energy neutrino experimental determination, including JUNO results on solar neutrino parameters \cite{JUNO:2025gmd}, and showing how constraints are sensitive to the experimental advance. 

\subsection{Seesaw mechanism} 

In the type-I seesaw extension of the SM, one introduces $N$ RH neutrinos $\nu_{RI}$ ($I=1,\dots,N$) with Yukawa couplings to 
lepton and Higgs doublets and a Majorana mass term. Inspired by $SO(10)$ models we consider the case $N=3$. 
The  leptonic mass terms generated after electroweak spontaneous symmetry breaking by the Higgs vacuum expectation value,
$v =174\,{\rm GeV}$, can be written in the flavour basis 
where both charged lepton and Majorana mass terms are diagonal
as ($\a=e,\m,\t$ and $I=1,2,3$)
\be
- {\cal L}^{m}_{{\ell}+\nu} = \,  \overline{\a_L} \, D_{m_{\ell}}\,\a_R + 
                              \overline{\nu_{\a L}}\,m_{D\a I} \, N_{R I} +
                               {1\over 2} \, \overline{N^{c}_{R I}} \, D_{M} \, N_{R I}  + \mbox{\rm h.c.}\,  ,
\ee
where $\widetilde{\Phi} \equiv {\rm i}\,\sigma_2\,\Phi^{\star}$ is the dual Higgs field. We also denoted by
$D_{m_{\ell}} \equiv {\rm diag}(m_e,m_{\m},m_{\t})$ the diagonal charged lepton mass matrix, 
by $D_{M}\equiv {\rm diag}(M_1,M_2,M_3)$ the diagonal Majorana mass matrix
and $m_{D}=h \, v$ is the neutrino Dirac mass matrix with $h$ the neutrino Yukawa matrix. 

In the seesaw limit, for $M \gg m_D$, the mass spectrum splits into two sets of Majorana neutrino eigenstates,
a light set, $\nu_{1'}$, $\nu_{2'}$ and $\nu_{3'}$, with masses $m_{1'} \leq m_{2'} \leq m_{3'}$ respectively, 
and a heavy set, $N_1$, $N_2$ and $N_3$, 
with masses almost coinciding with the three $M_I$ in $D_M$. The light neutrino masses are 
given by the seesaw formula \cite{seesaw}
\be\label{seesaw}
D_m =  U_\nu^{\dagger} \, m_D \, {1\over D_M} \, m_D^T  \, U_\nu^{\star}  \,   ,
\ee
with $D_m = {\rm diag}(m_{1'},m_{2'},m_{3'})$.  The matrix $U_\nu$, diagonalising the light neutrino mass
matrix $m_{\nu} = -m_D\,M^{-1}\,m_D^T$ in the flavour basis, can be identified
with the PMNS leptonic mixing matrix. 

\subsection{Low energy neutrino data}

Neutrino oscillation experiments measure mass squared differences. They cannot measure the absolute neutrino mass scale.
Moreover,  current neutrino oscillation experiments cannot solve an ambiguity between normally ordered neutrino masses (NO)
and inverted ordered neutrino masses (IO). In the first case we can identify $m_1 = m_{1'}$, $m_2 = m_{2'}$ and $m_3 = m_{3'}$.
Most recent global analyses find  for the solar neutrino mass scale \cite{nufit25}
\be
m_{\rm sol}\equiv \sqrt{m^{\, 2}_2 - m_1^{\, 2}} = (8.68^{+0.054}_{-0.058})\,{\rm meV} \,  ,
\ee
and for the atmospheric neutrino mass scale 
\be
m_{\rm atm}\equiv \sqrt{m^{\, 2}_3 - m_1^{\, 2}} = (50.11^{+0.21}_{-0.20})\,{\rm meV}  \,   .
\ee
The sum of the two scales yields a lower bound on the sum of the neutrino masses:
\be
\sum_i m_i \geq m_{\rm sol} + m_{\rm atm} = (58.79^{+0.20}_{-0.21})\, {\rm meV} \geq 58.4\,{\rm meV}  \;\; (95\% {\rm C.L.}) \,  .
\ee
The IO case corresponds to $m_{1'} = m_{3}$, $m_{2'} = m_{1}$ and $m_{3'} = m_{2}$.
In this case global analyses find  for the solar neutrino mass scale \cite{nufit25}
\be
m_{\rm sol}\equiv \sqrt{m^{\, 2}_2 - m_1^{\, 2}} = (8.68^{+0.054}_{-0.058})\,{\rm meV} \,  ,
\ee
and for the atmospheric neutrino mass scale 
\be
m_{\rm atm}\equiv \sqrt{m^{\, 2}_2 - m_3^{\, 2}} = (49.83^{+0.20}_{-0.20})\,{\rm meV}  \,   .
\ee
The sum of neutrino masses in the IO case is  much more stringently lower bounded:
\be
\sum_i m_i \geq m_{\rm atm} + \sqrt{m^2_{\rm atm}-m^2_{\rm sol}} = (98.94 \pm 0.40)\, {\rm meV} \geq  98.1\,{\rm meV}  \;\; (95\% {\rm C.L.}) \,  .
\ee
This is in tension with the cosmological upper bound  on the sum of neutrino masses obtained
combining CMB anisotropy, lensing and baryon acoustic oscillation data \cite{Tristram:2023haj}:
\be\label{upperbm1}
\sum_i m_i < 110\,{\rm meV} \;\; (95\% \, {\rm C.L.}) \,  ,
\ee
that, taking into account the values for the solar and atmospheric neutrino mass scales,  translate
into an upper bound on the lightest neutrino mass for NO (IO):
\be\label{upperbm1}
m_{1'} \lesssim  26 \, ( 9)\,{\rm meV} \,  . 
\ee
The leptonic mixing matrix can be parameterised in terms of the usual three mixing angles 
$\theta_{12}$, $\theta_{23}$ and $\theta_{13}$, the Dirac phase $\d$
and the Majorana phases $\rho$ and $\s$, as
\be\label{UPMNS}
U_\nu=  \left( \begin{array}{ccc}
c_{12}\,c_{13} & s_{12}\,c_{13} & s_{13}\,e^{-{\rm i}\,\d} \\
-s_{12}\,c_{23}-c_{12}\,s_{23}\,s_{13}\,e^{{\rm i}\,\d} &
c_{12}\,c_{23}-s_{12}\,s_{23}\,s_{13}\,e^{{\rm i}\,\d} & s_{23}\,c_{13} \\
s_{12}\,s_{23}-c_{12}\,c_{23}\,s_{13}\,e^{{\rm i}\,\d}
& -c_{12}\,s_{23}-s_{12}\,c_{23}\,s_{13}\,e^{{\rm i}\,\d}  &
c_{23}\,c_{13}
\end{array}\right)
\, {\rm diag}\left(e^{i\,\rho}, 1, e^{i\,\sigma}
\right)\,   ,
\ee
where $s_{ij}\equiv \sin\theta_{ij}$ and $c_{ij}\equiv \cos\theta_{ij}$. 
In the NO case latest global analyses of neutrino oscillation experiment results, including atmospheric neutrino
data from Super-Kamiokande and IceCube collaborations,  find for the mixing angles and 
leptonic Dirac phase $\d$ the following best fit values, $1\s$ errors  and $3\s$ intervals \cite{nufit25}: 
\bea\label{exprangesNO}
\theta_{13} & = &  8.62^{\circ}\pm 0.11^{\circ} \in [8.26^{\circ}, 8.95^{\circ}] \,  , \\ \nonumber
\theta_{12} & = &  {33.76^{\circ}}^{+0.42^\circ}_{-0.41^\circ} \in [32.54^{\circ}, 35.03^{\circ}]  \,  , \\ \nonumber
\theta_{23} & = &  {43.29^{\circ}}^{+0.96^{\circ}}_{-0.79^\circ} \in [41.27^{\circ}, 49.86^{\circ}]  \,  ,  \\ \nonumber
\d & = &  {-148^{\circ}} ^{+26^{\circ}}_{-36^{\circ}} \in  [-235^{\circ}, 5^{\circ}]  \, .
\eea 
As one can notice, there is a $3\s$ exclusion interval , $\d\ni [5^\circ, 125^{\circ}]$, 
for the Dirac phase that disfavours $\sin\d > 0$.  In the IO case one finds \cite{nufit25}:
\bea\label{exprangesIO}
\theta_{13} & = &  8.65^{\circ}\pm 0.11^{\circ} \in [8.32^{\circ}, 8.99^{\circ}] \,  , \\ \nonumber
\theta_{12} & = &  {33.76^{\circ}}^{+0.42^\circ}_{-0.41^\circ} \in [32.54^{\circ}, 35.03^{\circ}]  \,  , \\ \nonumber
\theta_{23} & = &  {47.90^{\circ}}^{+0.73^{\circ}}_{-0.92^\circ} \in [41.51^{\circ}, 49.83^{\circ}]  \,  ,  \\ \nonumber
\d & = &  {-86^{\circ}} ^{+22^{\circ}}_{-25^{\circ}} \in  [-157^{\circ}, -25^{\circ}]  \, .
\eea 
Also in this case there is  a $3\s$ exclusion interval , $\d\ni [-25^\circ, 203^{\circ}]$, 
for the Dirac phase that disfavours $\sin\d > 0$.  
The IO case is disfavoured at $\sim 2.4\s$ compared to NO.\footnote{This is still excluding the new JUNO results we
discussed in the introduction. Likely inclusion of those results would bring global analyses to disfavour IO even at $3\s$ or more.
A recent combined analysis of all atmospheric neutrino data plus 207 days of JUNO data excludes IO at $3.9\sigma$ \cite{Jin:2026ejo}.} 
However, since data cannot be taken as conclusive, we will still consider IO as a viable option.

No neutrinoless double beta ($0\nu\b\b$) decay signal has been detected so far.  This implies that
there are no experimental constraints on the Majorana phases and that experiments can only 
place an upper bound on the effective  $0\nu\b\b$ neutrino mass $m_{ee} \equiv |m_{\nu ee}|$. 
The most stringent one has been set by the KamLAND-Zen collaboration, that found \cite{KamLAND-Zen:2024eml}
\be\label{upperbmee}
m_{ee} \leq (28 \mbox{--} 122)\,{\rm meV}  \;\;   (90\%\, {\rm C.L.}) \,  ,
\ee
where the range accounts for nuclear matrix element uncertainties.  This translates into an upper bound 
on the lightest neutrino mass $m_{1'} \leq (84 \mbox{--} 353) \, {\rm meV}$ (90\%\, {\rm C.L.}). When this is compared with the upper bound Eq.~(\ref{upperbm1}) from cosmological observations (assuming $\Lambda$CDM), one can clearly see how the latter is much more stringent.\footnote{It is even more stringent
than it looks like, considering that the cosmological upper bound is at $95\%$ C.L. and the
upper bound from $0\nu\b\b$  at $90\%$ C.L. .}
However, it should be stressed that a  future $0\nu\b\b$ positive signal with a measurement of  $m_{ee}$ would not only provide 
independent information on the absolute neutrino mass scale but also some partial information on the Majorana phases, in addition of course to establish
Majorana nature of neutrinos and lepton number violation. The comparison with an independent determination of the absolute neutrino mass scale,
places important constraints within $SO(10)$-inspired leptogenesis \cite{riotto2,decrypting,DiBari:2020plh}. 

Finally, the KATRIN experiment has recently placed the upper bound 
\be\label{upperbmnue}
m_{\nu_e} \lesssim 0.45\,{\rm eV} \;\;\;  (90\% \,{\rm C.L.})
\ee 
on the effective electron neutrino mass \cite{Katrin:2024tvg}.
Since this falls in the quasi-degenerate limit,  it translates into an equal upper bound on $m_{1'}$.

\subsection{$SO(10)$-inspired conditions}

The neutrino Dirac mass matrix, written in the flavour basis, can be parameterised in the bi-unitary parameterisation 
(mathematically, the singular value decomposition) as
\be\label{svd}
m_D = V^\dagger_L\, D_{m_{D}} \, U_R \,   ,
\ee
where $D_{m_{D}} \equiv {\rm diag}(m_{D1},m_{D2},m_{D3})$.
The unitary matrices $V_L$ and $U_R$ transform, respectively, the LH and RH neutrino fields from the flavour basis to the Yukawa basis. 

We can now specify quantitatively the first $SO(10)$-inspired condition (I) 
discussed in the introduction. The three Dirac neutrino masses, 
$m_{Di}$ ($i=1,2,3$), can be expressed in terms of the three up-quark masses as:
\be\label{mDi}
m_{D1} = \alpha_1 \, m_{\rm up} , \; 
m_{D2} = \alpha_2 \, m_{\rm charm} , \;
m_{D3} = \alpha_3 \, m_{\rm top} , 
\;\; \mbox{\rm with} \;\; \alpha_i = {\cal O}(1) \; (i=1,2,3) .
\ee
For the three values of the up quark masses we adopt the three values at the scale
of the $N_2$ mass, where they are relevant, $M_2 \sim 10^{11}\,{\rm GeV}$, using the same reference
values used in \cite{riotto1} (taken from \cite{fusaoka}), namely: $m_{\rm up} = 1\,{\rm MeV}$,
$m_{\rm charm} =400\,{\rm MeV}$, $m_{\rm top} = 100\,{\rm GeV}$.  
These values define  $\alpha_1, \alpha_2$ and $\alpha_3$.
The condition in Eq.~(\ref{mDi}) implies hierarchical Dirac neutrino masses, 
$m_{D1} \ll m_{D2} \ll m_{D3}$. In this case one has
the following  analytical expressions for the heavy 
neutrino masses \cite{Akhmedov:2003dg,decrypting,full}
\be\label{RHspectrum}
M_1    \simeq    {m^2_{D1} \over |(\widetilde{m}_\nu)_{11}|} \, , \;\;
M_2  \simeq     {m^2_{D2}  \over m_1 \, m_2 \, m_3 } \, {|(\widetilde{m}_{\nu})_{11}| \over |(\widetilde{m}_{\nu}^{-1})_{33}|  } \,  ,  \;\;
M_3  \simeq   m^2_{D3}\,|(\widetilde{m}_{\nu}^{-1})_{33}| \, ,
\ee
and the RH neutrino mixing matrix
\be\label{RHnumix}
U_R \simeq  
\left( \begin{array}{ccc}
1 & -{m_{D1}\over m_{D2}} \,  {\widetilde{m}^\star_{\nu 1 2 }\over \widetilde{m}^\star_{\nu 11}}  & 
{m_{D1}\over m_{D3}}\,
{ (\widetilde{m}_{\n}^{-1})^{\star}_{13}\over (\widetilde{m}_{\n}^{-1})^{\star}_{33} }   \\
{m_{D1}\over m_{D2}} \,  {\widetilde{m}_{\nu 12}\over \widetilde{m}_{\nu 11}} & 1 & 
{m_{D2}\over m_{D3}}\, 
{(\widetilde{m}_{\n}^{-1})_{23}^{\star} \over (\widetilde{m}_{\n}^{-1})_{33}^{\star}}  \\
 {m_{D1}\over m_{D3}}\,{\widetilde{m}_{\nu 13}\over \widetilde{m}_{\nu 11}}  & 
- {m_{D2}\over m_{D3}}\, 
 {(\widetilde{m}_\nu^{-1})_{23}\over (\widetilde{m}_\nu^{-1})_{33}} 
  & 1 
\end{array}\right) 
\,  D_{\Phi} \,  ,
\ee
with the three phases in 
$D_{\phi} \equiv {\rm diag}(e^{-i \, {\Phi_1 \over 2}}, e^{-i{\Phi_2 \over 2}}, e^{-i{\Phi_3 \over 2}})$ 
given by \cite{full}
\bea
\Phi_1 & =  & {\rm Arg}[-\widetilde{m}_{\nu 11}^{\star}] \,  , \\
\Phi_2 & = & {\rm Arg}\left[{\widetilde{m}_{\nu 11}\over (\widetilde{m}_{\nu}^{-1})_{33}}\right] -2\,(\rho+\s) - 2\,(\rho_L + \s_L) \, , \\
\Phi_3 & =  & {\rm Arg}[-(\widetilde{m}_{\nu}^{-1})_{33}] \,  .
\eea
One can also derive an  expression for the 
orthogonal matrix. Starting from its definition
$\O = D_m^{-{1\over 2}}\, U^{\dagger} \, m_D \,  D_M^{-{1\over 2}} $ \cite{Casas:2001sr}
and using Eq.~(\ref{svd}), one obtains for its matrix elements ($\a=e,\m,\t$; $i=1,2,3$; $I=1,2,3$),\footnote{The different labels denote the three different
sets of lepton flavours. While the charged lepton neutrino flavours and the neutrino mass eigenstates form two orthonormal bases,
the heavy neutrino lepton flavours cannot in order for the $C\!P$ asymmetries not to all vanish and have successful leptogenesis.}
\be\label{Oij}
\O_{iJ} \simeq {1\over \sqrt{m_i \, M_J}} \, 
\sum_{l=1}^{3} \, m_{D l} \, U^{\star}_{\a i}\,V^{\star}_{L\,l \a}\,U_{R \, l J} \,  ,
\ee
and from this, plugging Eq.~(\ref{RHspectrum}) and Eq.~(\ref{RHnumix}), one finds \cite{full}
\be\label{Omegaapp}
\hspace{-9mm}\O \simeq 
\left( \begin{array}{ccc}
i\, {(\widetilde{m}_{\nu}\,W^{\star})_{11} \over \sqrt{m_1 \, \widetilde{m}_{\nu 11}}} & 
\sqrt{m_2\,m_3\,(\widetilde{m}_{\nu}^{-1})_{33} \over \widetilde{m}_{\nu 11}}\,
\left(W^{\star}_{2 1} - W^{\star}_{31}\,{{(\widetilde{m}_{\nu}^{-1})_{23}}\over (\widetilde{m}_{\nu}^{-1})_{33}}\right) & 
{W^{\star}_{31}\over \sqrt{m_1\,(\widetilde{m}_{\nu}^{-1})_{33}}} \\
i\, {(\widetilde{m}_{\nu}\,W^{\star})_{12} \over \sqrt{m_2 \, \widetilde{m}_{\nu 11}}} & 
\sqrt{m_1\,m_3\,(\widetilde{m}_{\nu}^{-1})_{33} \over \widetilde{m}_{\nu 11}}\,
\left(W^{\star}_{22} - W^{\star}_{32}\,{{(\widetilde{m}_{\nu}^{-1})_{23}}\over (\widetilde{m}_{\nu}^{-1})_{33}}\right)  
& {W^{\star}_{32}\over \sqrt{m_2\,(\widetilde{m}_{\nu}^{-1})_{33}}}  \\
i\, {(\widetilde{m}_{\nu}\,W^{\star})_{13} \over \sqrt{m_3 \, \widetilde{m}_{\nu 11}}}  & 
\sqrt{m_1\,m_2\,(\widetilde{m}_{\nu}^{-1})_{33} \over \widetilde{m}_{\nu 11}}\,
\left(W^{\star}_{2 3} - W^{\star}_{3 3}\,{{(\widetilde{m}_{\nu}^{-1})_{23}}\over (\widetilde{m}_{\nu}^{-1})_{33}}\right)  
& {W^{\star}_{33}\over \sqrt{m_3\,(\widetilde{m}_{\nu}^{-1})_{33}}}  
\end{array}\right)  \,   ,
\ee
where $W \equiv V_L \, U$. The validity of these expressions clearly breaks down in the close vicinity
of the two level crossings where either  $(\widetilde{m}_{\nu}^{-1})_{33}$, or $\widetilde{m}_{\nu 11}$,
or  both, vanish.  The compact spectrum scenario \cite{compact} corresponds to a situation close to the third case.  
In all three cases the moduli of the orthogonal matrix entries become very large, a clear indication of the fine tuning that is needed in 
the seesaw formula to satisfy neutrino oscillation experimental data. 

Let us now parametrise the matrix $V_L$ analogously to the 
leptonic mixing matrix parameterisation in Eq.~(\ref{UPMNS}), introducing three mixing angles 
$\theta_{12}^{L}, \theta_{23}^L, \theta_{13}^L$
and three phases $\rho_L, \sigma_L, \delta_L$. 
We can express quantitatively the $SO(10)$-inspired condition (II) discussed in the introduction as
\be
I \lesssim V_L \lesssim V_{\rm CKM} \,  ,
\ee
meaning that three mixing angles in $V_L$ should be in the range 
$0 \leq \theta_{ij}^L \lesssim \theta_{ij}^{\rm CKM}$. 
Explicitly, we adopted: $\theta_{12}^L \leq  13^{\circ} \simeq \theta_{12}^{\rm CKM} \equiv \theta_c$, 
$\theta_{23}^L \leq  2.4^{\circ}\simeq \theta_{23}^{\rm CKM}$, $\theta_{13}^L \leq  0.2^{\circ}\simeq \theta_{13}^{\rm CKM}$.

In the top panels of Fig.~1 and Fig.~2, for NO and IO respectively,
we show  scatter plots\footnote{In the case of NO, the scatter plot for $V_L=I$ contains about 50 million points. Only 500 hundred
points are found satisfying $M_3/M_2 \lesssim 2$, only $7$ for  $M_2/M_1 \lesssim 2$ and none for $M_1 \simeq M_2 \simeq M_3$
(for these 500 point masses are calculated numerically, i.e., not using Eq.~(\ref{RHspectrum})). 
Similar results are obtained for $I \lesssim V_L \lesssim V_{\rm CKM}$. This
shows how extremely fine-tuned points close to crossing level solutions are. For IO they are even more rare.} for the three heavy neutrino masses that are obtained randomly scanning in the following way: 
\begin{itemize}
\item Uniformly on the six parameters in the matrix $V_L$: in the left panels for
$V_L = I$ and in the right panels for $I < V_L < V_{\rm CKM}$;
\item Uniformly logarithmically on the lightest neutrino mass $m_1$ in the range
$10^{-5}\,{\rm eV}$--$1\,{\rm eV}$; 
\item Uniformly on the three phases in $U$; 
\item Uniformly on the three mixing angles over approximately the 
following $5\sigma$ ranges:
\be
8^\circ \leq \theta_{13} \leq 9.2^\circ \,  , \;
31.5^\circ \leq \theta_{12} \leq  36^\circ \,  , \;
39^\circ \leq \theta_{23} \leq  51^\circ \,  ,
\ee
and selecting only points respecting the condition
\be
\chi^2 = 
\left({\theta_{12} - \bar{\theta}_{12} \over \sigma_{\theta_{12}}} \right)^2 +
\left({\theta_{13} - \bar{\theta}_{13} \over \sigma_{\theta_{13}}} \right)^2 
< \chi^2_{\rm max} = 25 \,  .
\ee
\end{itemize}
The mean values and standard deviations are those given in Eqs.~(\ref{exprangesNO})
and (\ref{exprangesIO}) for NO and IO, respectively. 
We do not include $\theta_{23}$ in the $\chi^2$, since its 
experimental error is still quite far from a Gaussian distribution. 
In this way all our scatter plots do not assume any {\em a priori}  
experimental preference for the first or second octant. 
\begin{figure}[t]
\psfig{file=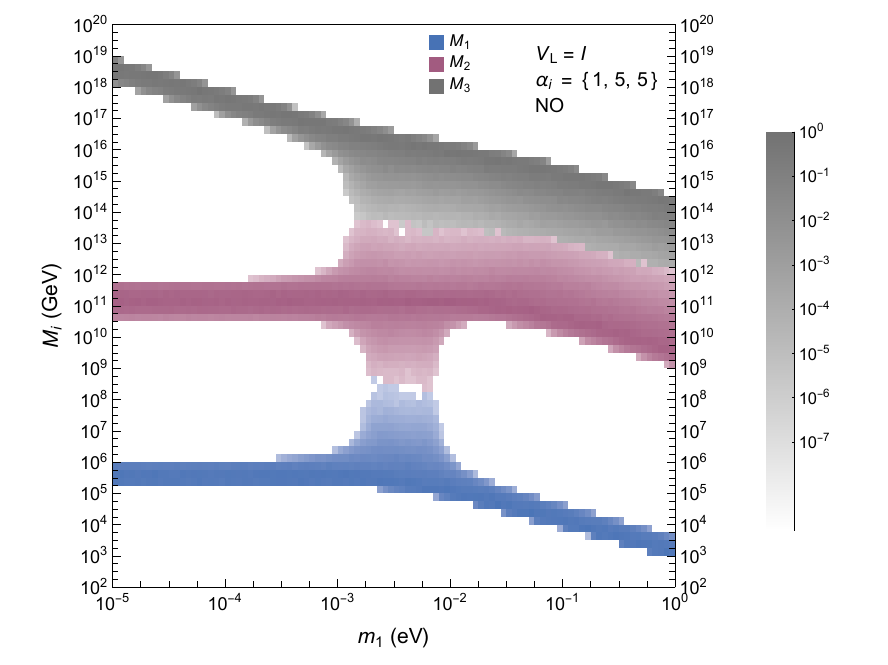,scale=0.5} \hspace{10mm}
\psfig{file=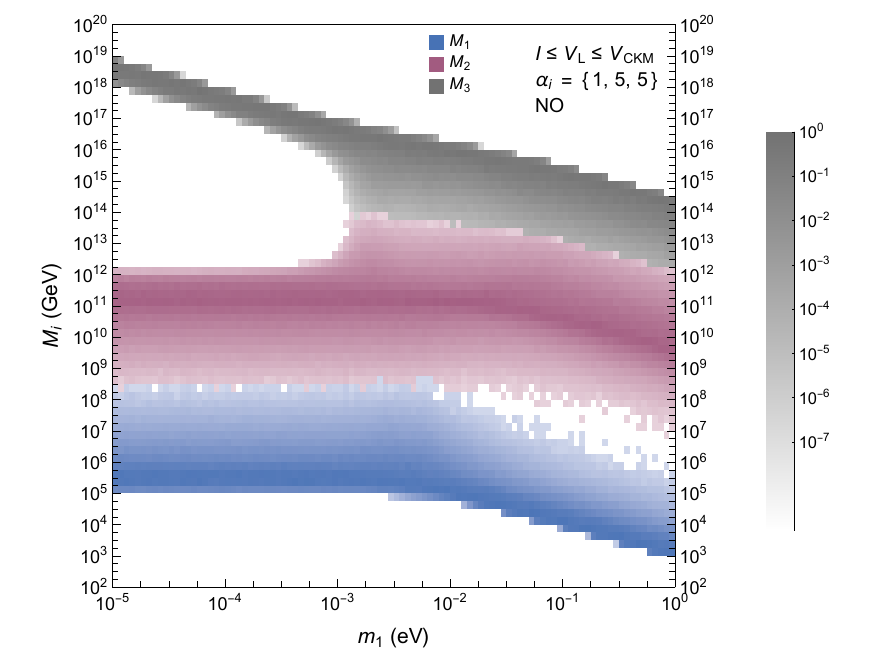,scale=0.5} \\
\psfig{file=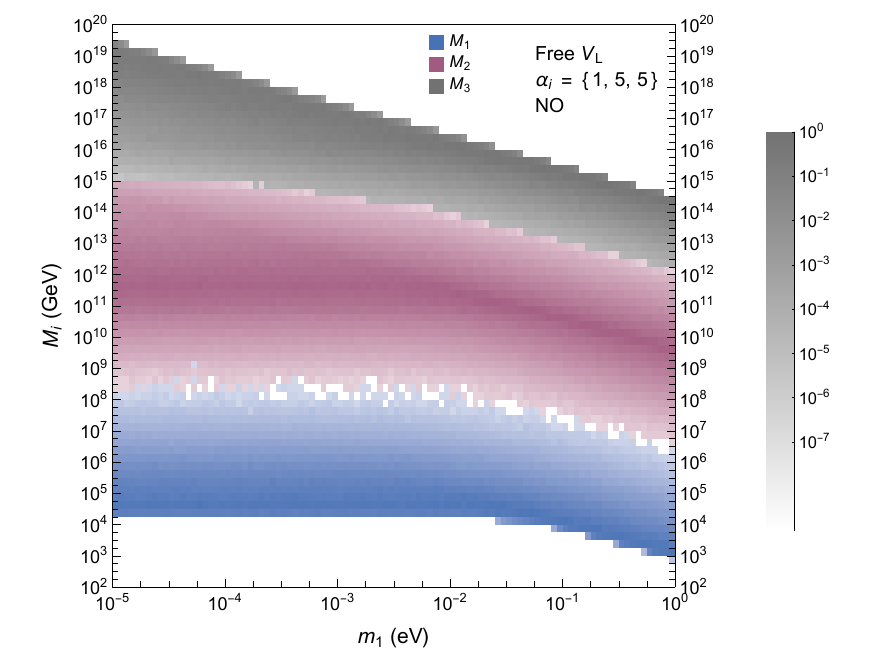,scale=0.5} \hspace{10mm}
\psfig{file=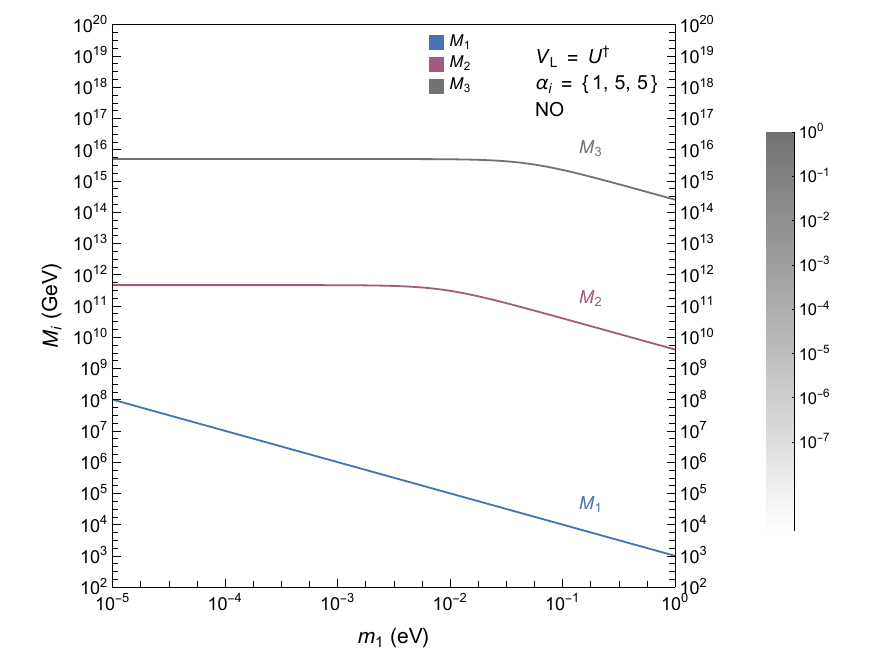,scale=0.5
}
\caption{Scatter plots showing viable solutions for the three heavy neutrino masses
obtained using the expressions in Eq.~(\ref{RHspectrum}) and scanning
over the parameters in $V_L$, $U$ and $m_1$, for NO, as described in the main text. 
No successful leptogenesis condition was  imposed. 
}
\end{figure}
Also notice that we are not including $m_{\rm sol}$ and $m_{\rm atm}$ in the calculation of the
$\chi^2$ since, due to the small experimental errors, these quantities can be 
fixed to their best fit values with no impact on all the results we derive. 
\begin{figure}[t]
\psfig{file=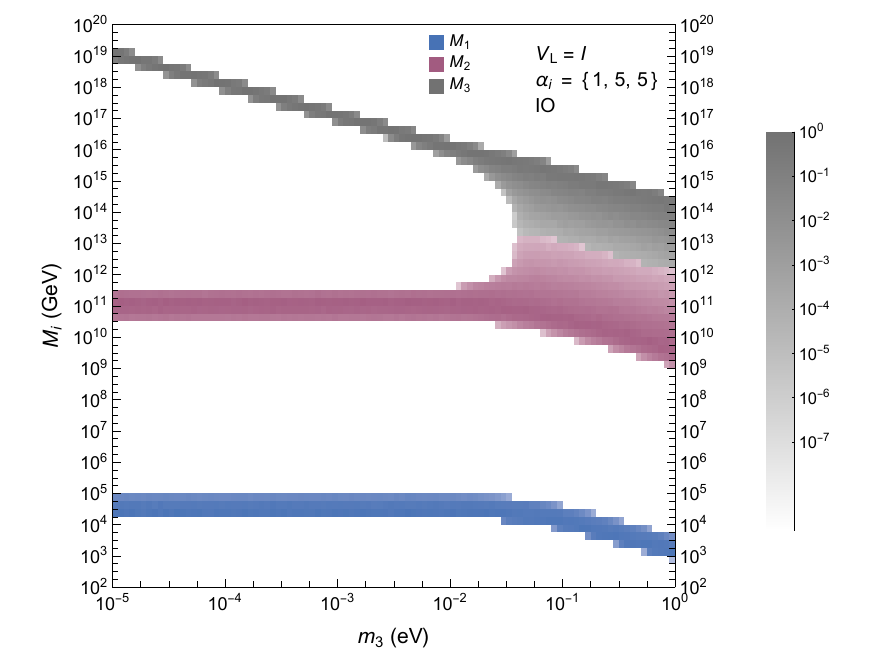,scale=0.5} \hspace{10mm}
\psfig{file=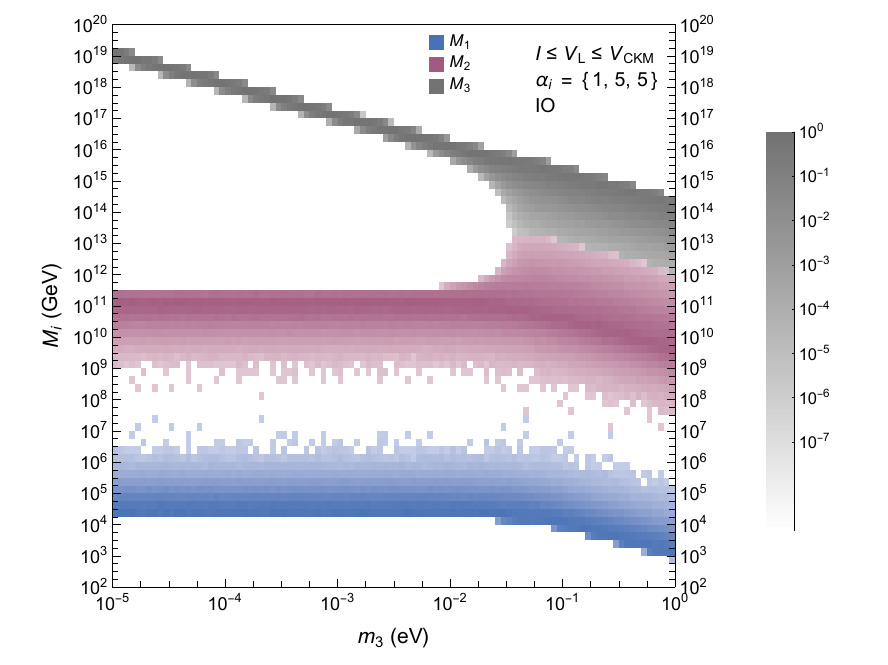,scale=0.5} \\
\psfig{file=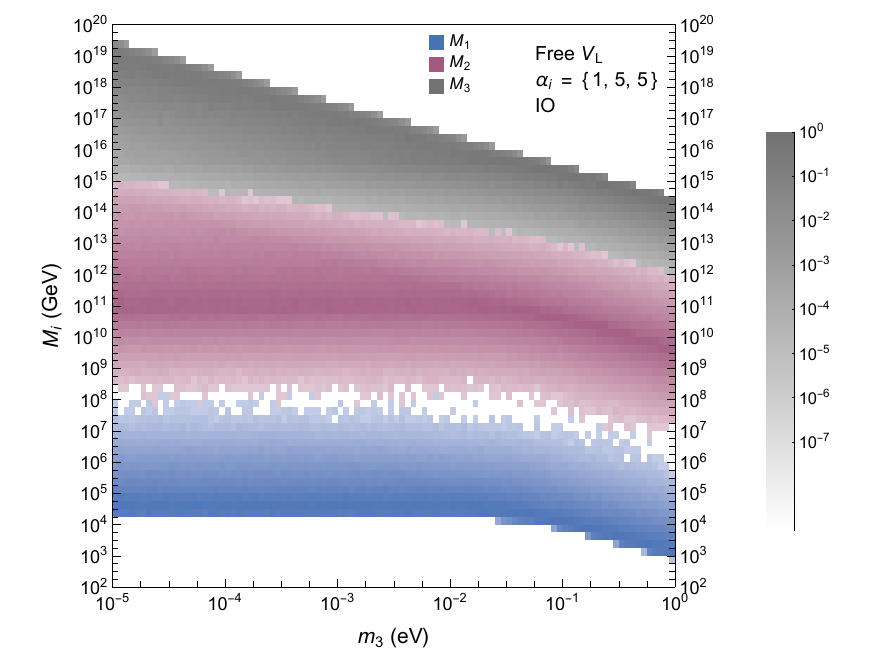,scale=0.5} \hspace{10mm}
\psfig{file=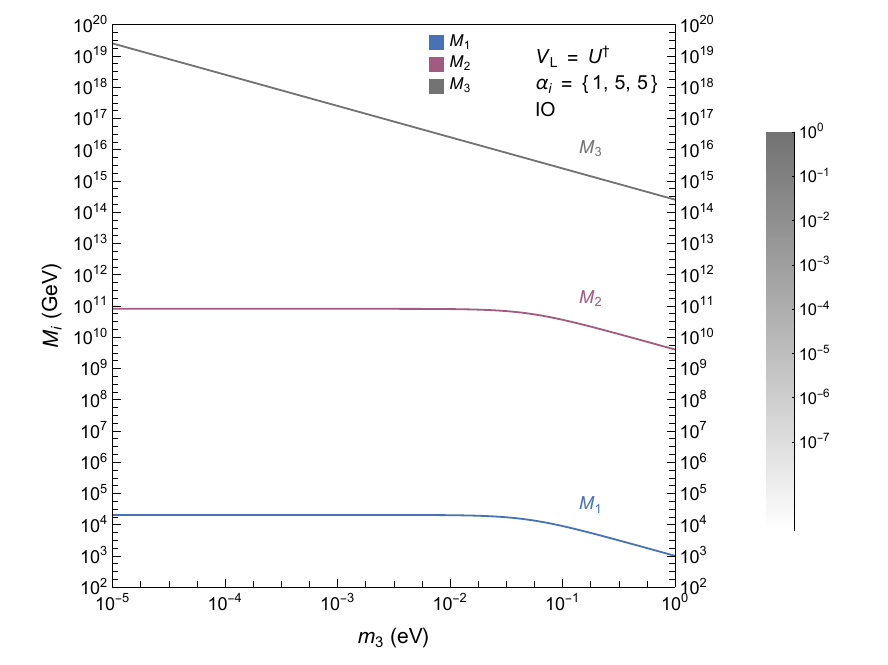,scale=0.5}
\caption{Same scatter plots as in Fig.~1 but for IO.}
\end{figure}
For a comparison, in the two top left panels we show also the results obtained if one imposes $V_L = I$.
This assumption is not justified theoretically but the comparison between left 
and right panels nicely show how turning on angles in $V_L$ broadens the
RH neutrino mass spectrum but still preserving the result 
$100\,{\rm GeV} \ll M_1 \ll 10^{9}\,{\rm GeV}$. This is an important feature of these scatter plots, since they
clearly pin down $N_2$-leptogenesis as the necessary scenario within which to search for successful leptogenesis
solutions.

\subsection{$SO(10)$-inspired leptogenesis}

In the $N_2$-leptogenesis scenario, taking into account production by $N_2$-decays and washout by $N_1$-inverse 
processes, but neglecting flavour coupling originating from spectator processes,  the final $B-L$ asymmetry can be calculated 
as the sum of the three contributions, one from each (charged lepton) flavour:
\cite{vives,bounds,fuller,density}:
\bea\label{twofl} \nonumber
N_{B-L}^{\rm lep, f} & \simeq &
\left[{K_{2e}\over K_{2\tau_2^{\bot}}}\,\ve_{2 \tau_2^{\bot}}\kappa(K_{2 \tau_2^{\bot}}) 
+ \left(\ve_{2e} - {K_{2e}\over K_{2\tau_2^{\bot}}}\, \ve_{2 \tau_2^{\bot}} \right)\,\kappa(K_{2 \tau_2^{\bot}}/2)\right]\,
\, e^{-{3\pi\over 8}\,K_{1 e}}+ \\ \nonumber
& + &\left[{K_{2\mu}\over K_{2 \tau_2^{\bot}}}\,
\ve_{2 \tau_2^{\bot}}\,\kappa(K_{2 \tau_2^{\bot}}) +
\left(\ve_{2\mu} - {K_{2\mu}\over K_{2\tau_2^{\bot}}}\, \ve_{2 \tau_2^{\bot}} \right)\,
\kappa(K_{2 \tau_2^{\bot}}/2) \right]
\, e^{-{3\pi\over 8}\,K_{1 \mu}}+ \\
& + &\ve_{2 \tau}\,\kappa(K_{2 \tau})\,e^{-{3\pi\over 8}\,K_{1 \tau}} \,  .
\eea
As one can see, there are three ingredients needed for an explicit calculation:
\begin{itemize}
\item The flavoured decay parameters
\be
K_{I\a} \equiv {\G_{I\a}+\overline{\G}_{I\a}\over H(T=M_I)}= 
{|m_{D\a I}|^2 \over M_I \, m_{\star}} \,  ,
\ee
where  $\Gamma_{I\a}=\Gamma (N_I \ra \phi^\dagger \, l_\alpha)$ 
and $\bar{\Gamma}_{I \a}=\Gamma (N_I \ra \phi \, \bar{l}_\alpha)$ are the
zero temperature limit of the flavoured decay rates into $\a$ leptons
and anti-leptons in the three-flavoured regime, $m_{\star}\equiv 16\,\pi^{5/2}\, \sqrt{g_{\star}}/(3 \sqrt{5})\,(v^2 / M_{\rm Pl}) \simeq 1.08 \,{\rm meV}$
is the equilibrium neutrino mass, $H(T)=\sqrt{g^{SM}_{\star}\,8\,\pi^3/90}\,T^2/M_{\rm P}$ is the expansion rate
and $g_{\star}^{SM}=106.75$ is the number of ultra-relativistic degrees of freedom in the standard model.
In Eq.~(\ref{twofl}) we have defined $K_{2\tau_2^{\bot}}\equiv K_{2e}+K_{2\mu}$.

Using the bi-unitary parameterisation Eq.~(\ref{svd}) for $m_D$,  
the flavoured decay parameters can be expressed as
\be\label{KialVL}
K_{I\a} = {\sum_{k,l} \, 
m_{Dk}\, m_{Dl} \,V_{L k\a} \, V_{L l \a}^{\star} \, U^{\star}_{R kI} \, U_{R l I} 
\over M_I \, m_{\star}}\,  .
\ee

\item The flavoured $C\!P$ asymmetries are defined by
\be
\ve_{2\a} \equiv -(\G_{2\a}-\overline{\G}_{2\a})/(\G_2 + \overline{\G}_2) \,  ,
\ee
where $\G_2 \equiv \sum_{\a} \G_{2\a}$ and $\overline{\G}_{2}\equiv \sum_{\a} \, \overline{\G}_{2\a}$.
They can be calculated using \cite{crv}
\be\label{eps2a}
\ve_{2\a} \simeq
\overline{\ve}(M_2) \, \left\{ {\cal I}_{23}^{\a}\,\x(M^2_3/M^2_2)+
\,{\cal J}_{23}^{\a} \, \frac{2}{3(1-M^2_2/M^2_3)}\right\}\, ,
\ee
where
\be
\overline{\ve}(M_2) \equiv {3\over 16\,\pi}\,{M_2\,m_{\rm atm} \over v^2} \, , 
\hspace{3mm} \xi(x)=\frac{2}{3}x\left[(1+x)\ln\left(\frac{1+x}{x}\right)-\frac{2-x}{1-x}\right] \,  ,
\ee
\be
{\cal I}_{23}^{\a} \equiv   {{\rm Im}\left[m_{D\a 2}^{\star}
m_{D\a 3}(m_D^{\dag}\, m_D)_{2 3}\right]\over M_2\,M_3\,\mtt\,m_{\rm atm} }\,   
\hspace{5mm}
\mbox{\rm and}
\hspace{5mm}
{\cal J}_{23}^{\a} \equiv  
{{\rm Im}\left[m_{D\a 2}^{\star}\, m_{D\a 3}(m_D^{\dag}\, m_D)_{3 2}\right] 
\over M_2\,M_3\,\mtt\,m_{\rm atm} } \,{M_2\over M_3}   ,
\ee
with $\mtt \equiv (m_D^{\dag}\, m_D)_{2 2}/M_2$.  However, the first term is important only when RH neutrino masses are quasi-degenerate,
close to crossing level solutions.  In our case we will impose always a hierarchical condition $M_3 > 2 \, M_2$ and $M_2  > 2\,M_1$\footnote{Potential solutions
with $M_3 \simeq M_2 \gtrsim 10^{12}\,{\rm GeV}$ are possible, though certainly very fine-tuned in different respects, 
but they would require an extension of our calculation with the use of density matrix equations
and accounting for asymmetry production and wash-out from the $N_3$'s.} 
and one can reliably use the approximate expression \cite{full}
\be\label{ve2a}
\ve_{2\a} \simeq {3 \over 16\, \pi}\, {M_2 \over M_3}   {{\rm Im}\left[m_{D\a 2}^{\star}
m_{D\a 3}(m_D^{\dag}\, m_D)_{2 3}\right]\over v^2 \, (m^\dagger_D\,m_D)_{22} }\,   .
\ee
Again, using the bi-unitary parameterisation, one obtains:
\be\label{ve2al}
\ve_{2\a} \simeq {3 \over 16\, \pi \, v^2}\,
{|(\widetilde{m}_{\nu})_{11}| \over m_1 \, m_2 \, m_3}\,
{\sum_{k,l} \,m_{D k} \, m_{Dl}  \,  {\rm Im}[V_{L k \a }  \,  V^{\star}_{L  l \a } \, 
U^{\star}_{R k 2}\, U_{R l 3} \,U^{\star}_{R 3 2}\,U_{R 3 3}] 
\over |(\widetilde{m}_{\nu}^{-1})_{33}|^{2} + |(\widetilde{m}_{\nu}^{-1})_{23}|^{2}}   \,  .
\ee
\item The efficiency factor at the production $\k(K_{2\a})$ can be calculated using 
a simple analytic expression valid for initial thermal abundance \cite{flavourlep}
\be\label{kappa}
\k(K_{2\a}) = 
{2\over z_B(K_{2\a})\,K_{2\a}}
\left(1-e^{-{K_{2\a}\,z_B(K_{2\a})\over 2}}\right) \,  , \;\; z_B(K_{2\a}) \simeq 
2+4\,K_{2\a}^{0.13}\,e^{-{2.5\over K_{2\a}}} \,   .
\ee
Since  all solutions are characterised by strong wash-out at the production
(either $K_{2\t} \gg 1$ or $K_{2 \tau_2^{\bot}} \gg 1$ respectively for tauon and muon-dominated solutions),
the final asymmetry does not depend on the initial $N_2$ abundance anyway. There are tauon-dominated solutions ($\tau_B$ solutions)
for which one can find values of $K_{2\tau}$ as low as $K_{2\tau}\simeq 1$. 
However, solutions with $K_{2\tau} \lesssim 3$, where there is some
dependence on the initial conditions and larger theoretical uncertainties, 
are realised only for values $m_1 \gtrsim 30\,{\rm meV}$, now disfavoured by the cosmological 
upper bound Eq.~(\ref{upperbm1}).  In any case, even starting from vanishing initial $N_2$-abundance,
there is no change on low energy neutrino parameters constraints. 
\end{itemize}
The expression in Eq.~(\ref{twofl}) is derived from momentum-integrated Boltzmann equations  
assuming a two-flavour regime production \cite{Barbieri:1999ma,flavoureffects,flavourlep}. 
It is strictly valid for masses $M_2\ll 10^{12}\,{\rm GeV}$ \cite{Blanchet:2006ch,DeSimone:2006nrs}.
For higher values, the production would make a transition  to a one-flavour regime where the wash-out would be stronger and one 
expects some suppression. In \cite{STSO10} solutions with $M_2 > 10^{12}\,{\rm GeV}$ were included calculating the asymmetry using a
one-flavour regime approximation. It was found that there were just some additional marginal solutions. Here
we just simply cut these marginal solutions for $M_2 > 10^{12}\,{\rm GeV}$ since in order to 
smoothly describe this transition getting the whole set of solutions,
density matrix equation should be solved  \cite{DeSimone:2006nrs,density}. 
In any case one can expect just marginal solutions appearing 
above $10^{12}\,{\rm GeV}$ and some below $10^{12}\,{\rm GeV}$ disappearing. 

From the final $B-L$ asymmetry one can calculate the baryon-to-photon number ratio \cite{Buchmuller:2004nz}
\be\label{etaBlep}
\eta_B^{\rm lep} =a_{\rm sph}\,{N_{B-L}^{\rm lep,f}\over N_{\g}^{\rm rec}} \simeq 
0.96\times 10^{-2}\,N_{B-L}^{\rm lep,f} \,  .
\ee
This expression accounts for sphaleron conversion \cite{manton,kuzmin,klebnikov,Harvey:1990qw} and standard photon dilution.
The second numerical expression holds when the abundances are normalised in a portion of comoving volume such that in ultra-relativistic 
thermal equilibrium the abundance of a RH neutrino $N_{N_i}^{\rm eq}(T \gg M_i) =1$. 
Successful leptogenesis requires $\eta_B^{\rm lep} = \eta_B^{\rm exp}$,
where $\eta_B^{\rm exp}$ is the measured value. Combining CMB plus BAO data, the {\em Planck} 
collaboration finds \cite{Tristram:2023haj}
\be\label{etaBexp}
\eta_{B}^{\rm exp} = (6.096 \pm 0.032)\, \times 10^{-10}  \,   .
\ee
The baryon-to-photon number ratio calculated from $SO(10)$-inspired leptogenesis is a 
function of the nine low energy neutrino parameters in $m_\nu$, the six parameters in $V_L$ and the parameter $\alpha_2$.
The latter can be conveniently fixed to a reasonable benchmark maximum value $\alpha_2^{\rm max}$, 
typically  $\alpha_2 =5$ \cite{riotto1,riotto2}. 
This means that the successful leptogenesis condition is equivalent to a parametric equation identifying
a hypersurface in the space of  parameters, explicitly:
\be\label{hypersurface}
\eta_B^{\rm lep}(m_{\rm sol},m_{\rm atm},m_1,\theta_{12},\theta_{13},\theta_{23},\delta,\rho,\sigma;
\theta_{12}^{\rm L},\theta_{13}^{\rm L},\theta_{23}^{\rm L},\delta^{\rm L},\rho^{\rm L},\sigma^{\rm L}; 
\alpha_2 \leq \alpha_2^{\rm max}) = \eta_B^{\rm exp} \,  .
\ee
Since points with $\eta_B > \eta_B^{\rm lep}$, i.e., over-exceeding the observed matter-antimatter asymmetry, 
can be easily reconciled with the successful leptogenesis condition for some value $\alpha_2 < \alpha_2^{\rm max}$,
we can simply impose $\eta_B > \overline{\eta}_B^{\rm exp} - 3\sigma = 6.000 \times 10^{-10}$.\footnote{Since we fixed $\chi^2_{\rm max}=25$,
we might also consistently take a $5\sigma$ lower bound for $\eta_B^{\rm lep}$. However, because of the small
error on $\eta_B$ there is actually no real difference (and we like a round value $6.000 \times 10^{-10}$). 
The same if we would decrease $\chi^2_{\rm max}$ from $25$ to $9$, but in this case solutions would be found less efficiently. 
The numerical determination of the hypersurface Eq.~(\ref{hypersurface}) is stable to such slight changes.}

In the left panel of Fig.~3  we show for NO scatter plots   
of the solutions respecting both low energy neutrino data and successful leptogenesis condition projected
on the three-dimensional space $(m_1,\delta,\theta_{23})$.  The asymmetry is calculated using Eq.~(\ref{twofl}) neglecting
flavour coupling.
\begin{figure}[t]
\begin{center}
\psfig{file=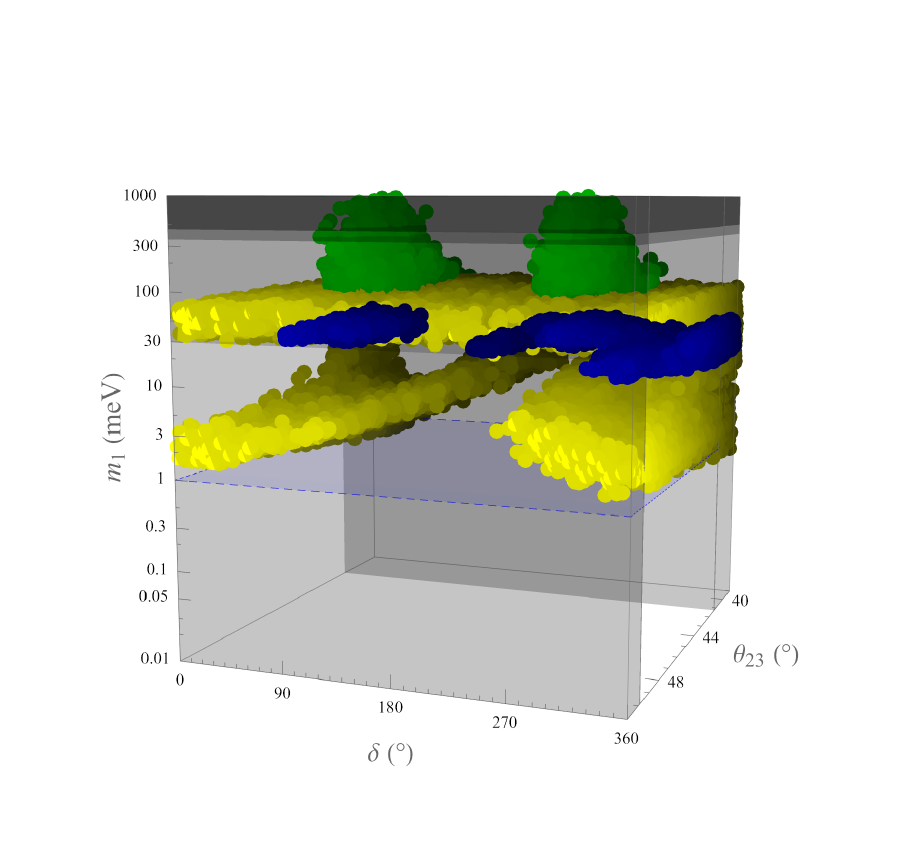,scale=0.52} \hspace{-10mm}
\psfig{file=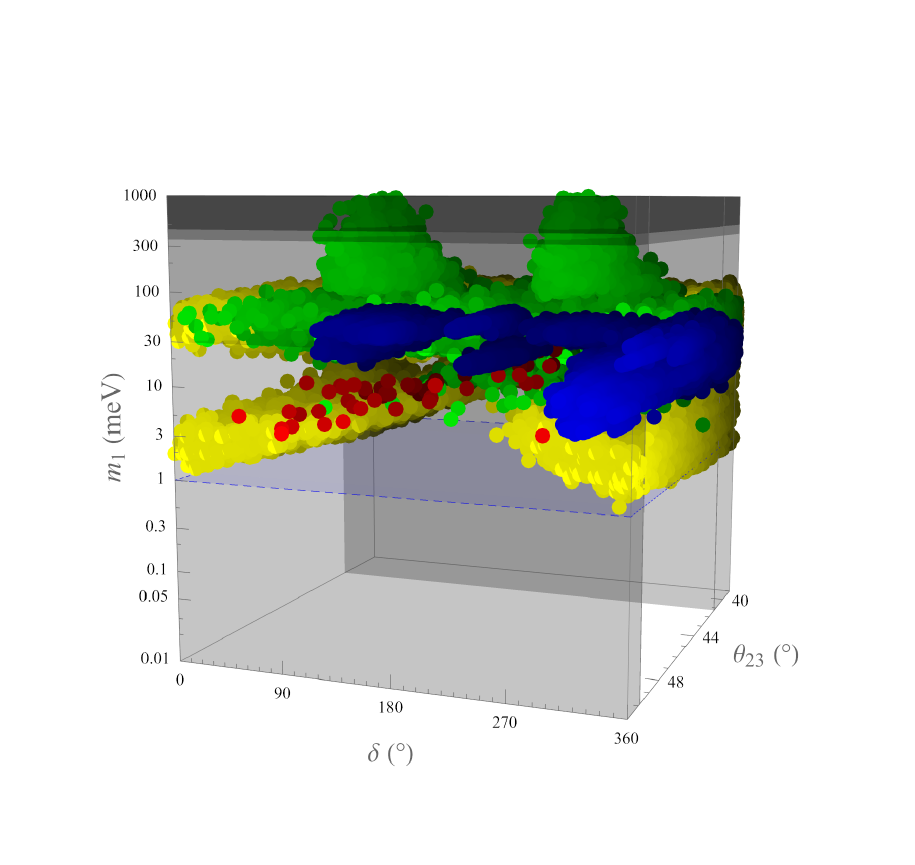,scale=0.52} 
\end{center}
\vspace{-5mm}
\caption{Scatter plots of the solutions obtained for NO imposing successful $SO(10)$-inspired leptogenesis,
projected in the three-dimensional space $(m_1,\delta,\theta_{23})$. The grey regions denote the $95\%$ excluded
values for $m_1$ and the $3\s$ excluded values for $\theta_{23}$. The dashed contoured planes highlight the lower bound on $m_1$.
The asymmetry is calculated neglecting and including flavour coupling in the left and right panel, respectively.
The colours indicate the flavour dominance (yellow, green and red for tauon, muon and electron-dominated solutions, respectively)
and strong thermal solutions (blue).}
\end{figure}
These are obtained for $I\leq V_L \leq V_{\rm CKM}$  from a random scan over the same ranges we used to generate the panels of 
Fig.~1 and Fig.~2 and for $\alpha_2^{\rm max} =5$ (but we use $\alpha_3=1$, in any case results are insensitive to the
value of $\alpha_3$). This is a benchmark reasonable maximum value
adopted also in all previous analyses in order to allow for a meaningful comparison. 
For NO, the results we obtain confirm once more the results obtained in previous works \cite{riotto1,riotto2}. In particular,
comparing with those we obtained in \cite{DiBari:2025zlv}, we can conclude that 
the small update in the $\theta_{12}$ experimental value in Eq.~(\ref{exprangesNO})
does not produce any change. 
\begin{figure}[t]
\begin{center}
\psfig{file=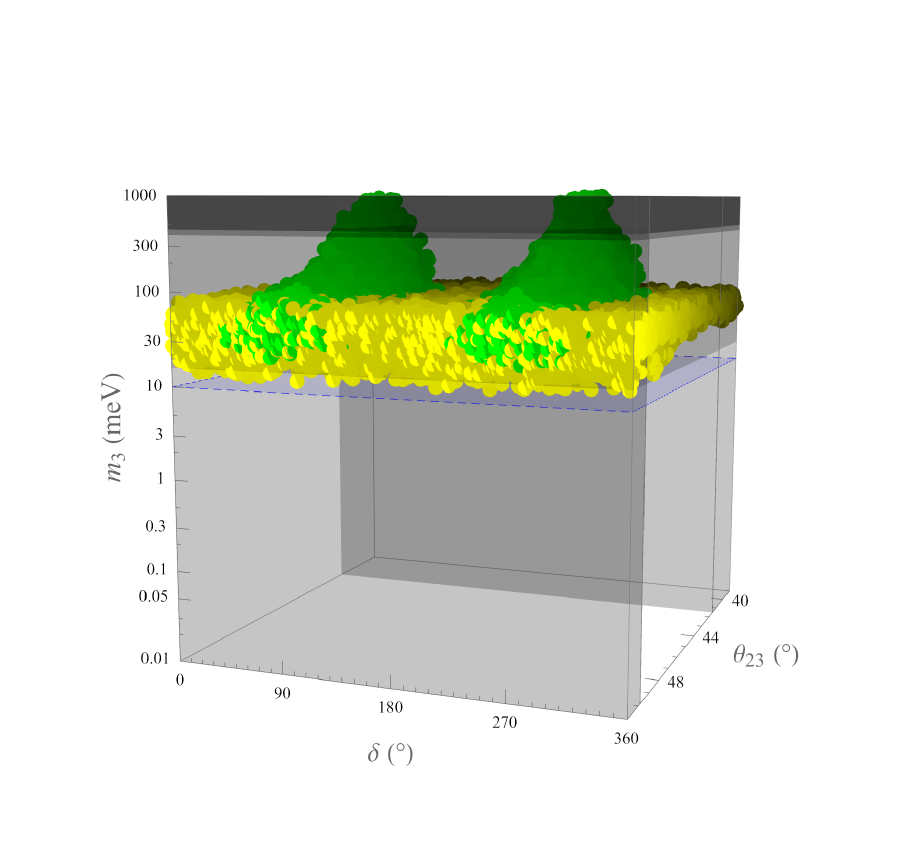,scale=0.5} \hspace{-10mm}
\psfig{file=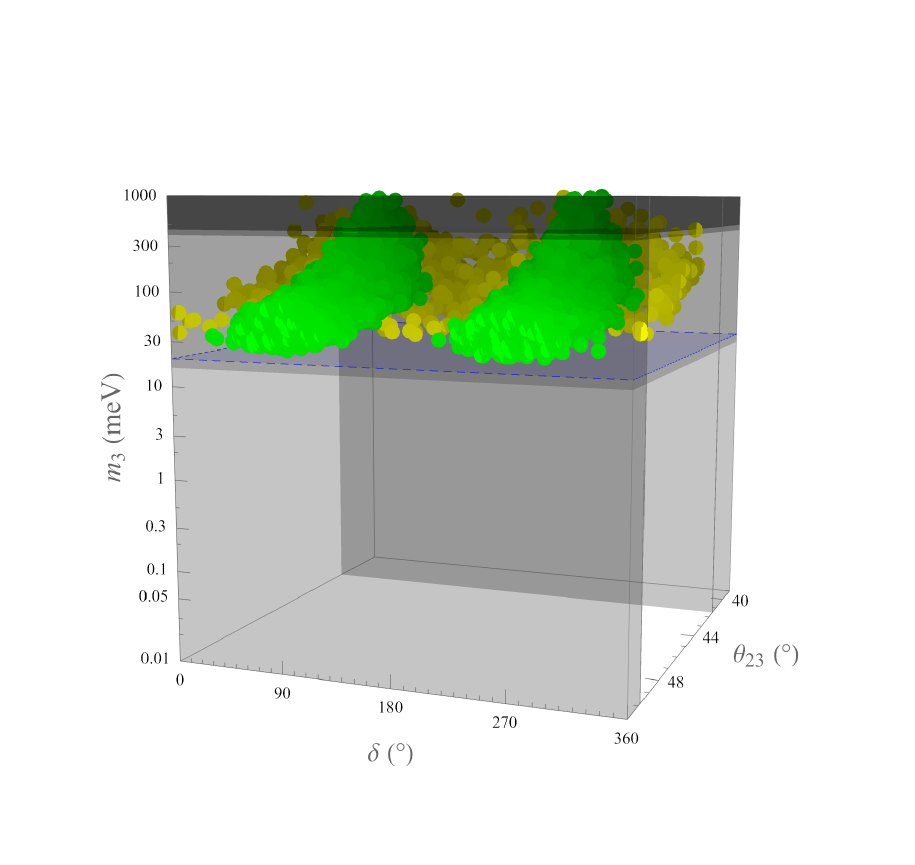,scale=0.5} 
\end{center}
\vspace{-10mm}
\caption{Same scatter plots as in Fig.~3 but for IO.}
\end{figure}
In Fig.~4 we show the same results for IO. These confirm (and strengthen) the conclusion reached in \cite{riotto2}: 
there are no solutions for IO compatible with current low energy neutrino data, in particular with the cosmological 
upper bound on neutrino masses combined with the upper bound on $\theta_{23}$.

In the left panel of Fig.~5 and Fig.~6, for NO and IO respectively, 
we also show the projections of the scatter plots onto the plane $m_1$--$\theta_{23}$. 
\begin{figure}[t]
\begin{center}
\psfig{file=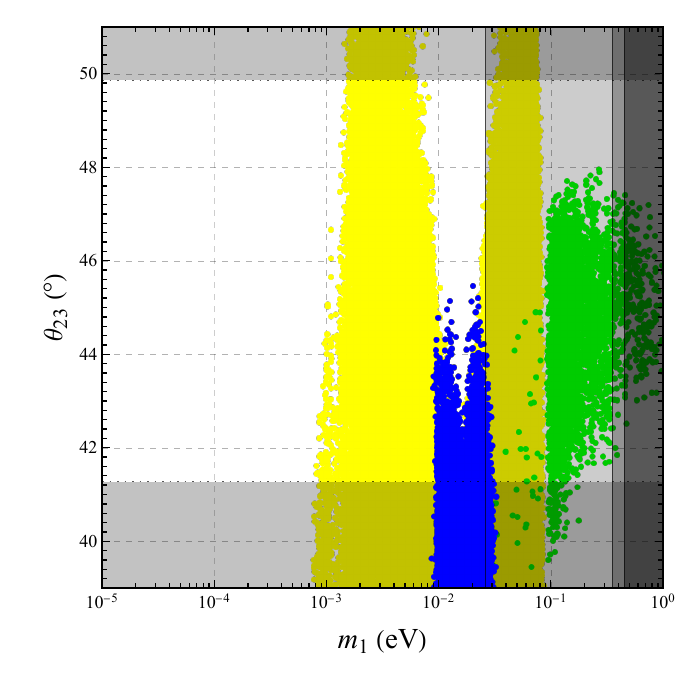,scale=0.5} \hspace{-0mm}
\psfig{file=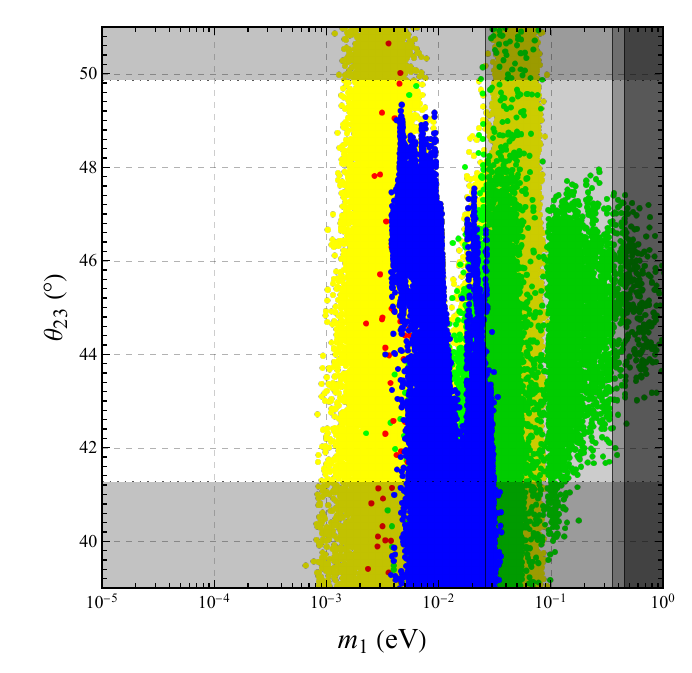,scale=0.5} 
\end{center}
\vspace{-8mm}
\caption{Projection of the scatter plot for NO shown in Fig.~3 onto the $m_1$--$\theta_{23}$ plane.}
\end{figure}
In the case of NO one can notice the interesting upper bound $\theta_{23} \lesssim 45^\circ$ for $10\,{\rm meV} \lesssim m_1 \lesssim 30\,{\rm meV}$.
In Fig.~6 this projection shows even more clearly how IO is incompatible with current cosmological upper bound on $m_1$ in Eq.~(\ref{upperbm1})
combined with the upper bound on $\theta_{23}$.
\begin{figure}[t]
\begin{center}
\psfig{file=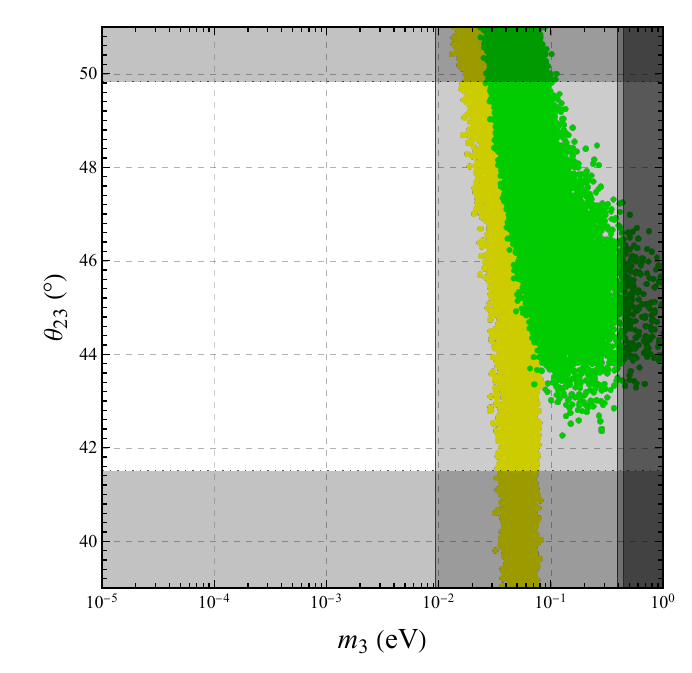,scale=0.5} \hspace{-0mm}
\psfig{file=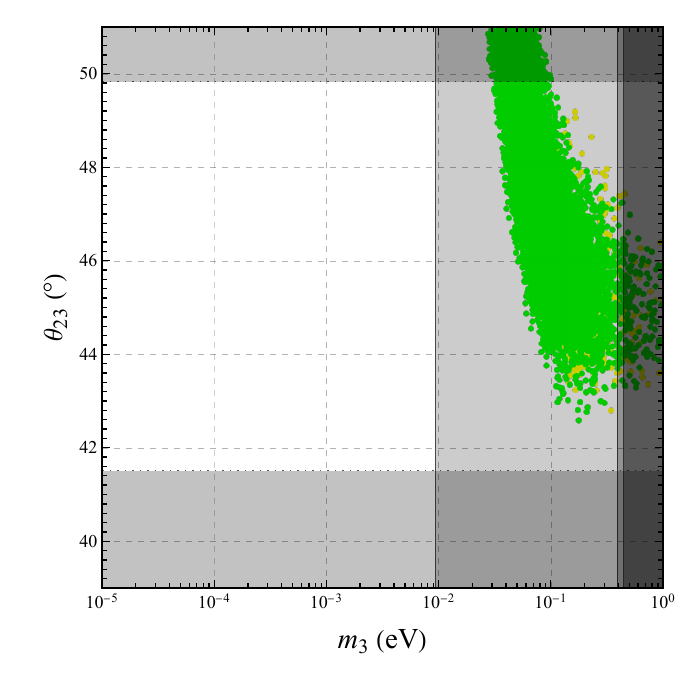,scale=0.5} 
\end{center}
\vspace{-8mm}
\caption{Projection of the scatter plot for IO shown in Fig.~4 onto the $m_1$--$\theta_{23}$ plane.}
\end{figure}

In the left panels of Fig.~7 and Fig.~8 we show the the projections of the scatter plots onto the plane $m_1$--$m_{ee}$ for NO and IO respectively.
\begin{figure}[t]
\begin{center}
\psfig{file=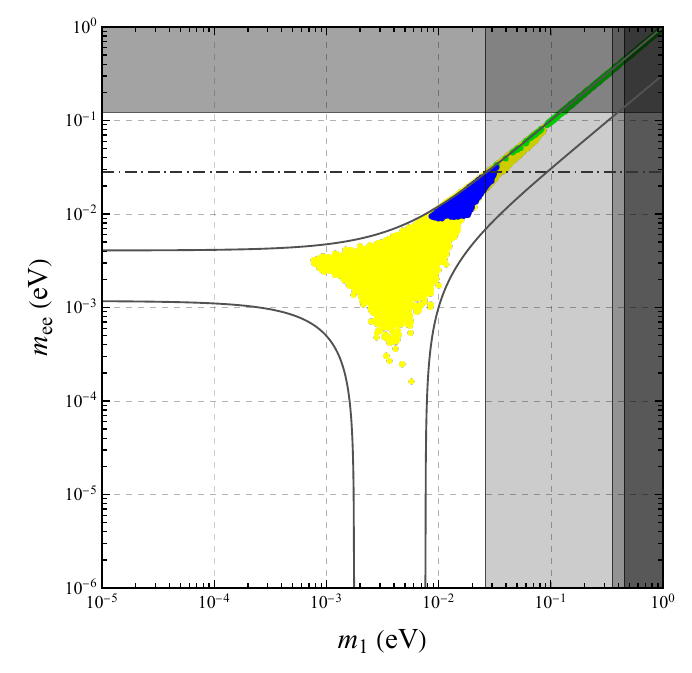,scale=0.5} \hspace{-0mm}
\psfig{file=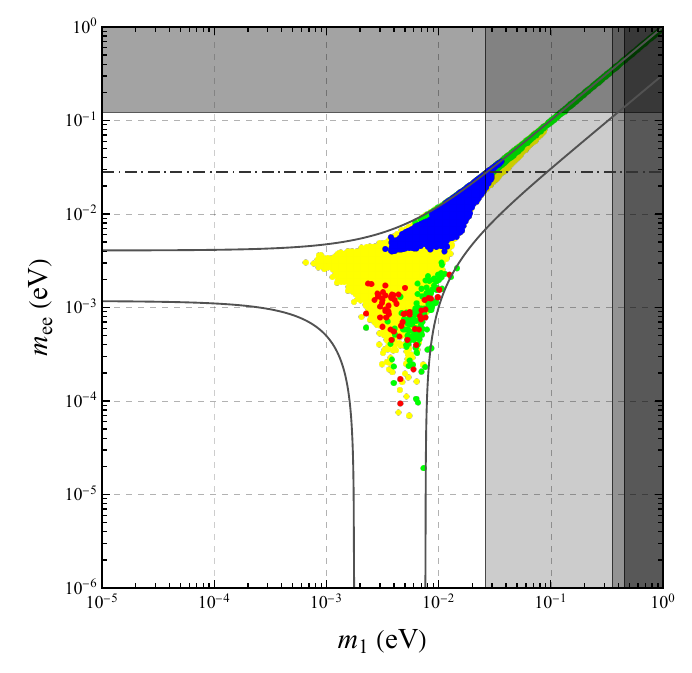,scale=0.5} 
\end{center}
\vspace{-8mm}
\caption{Projection of the scatter plot for NO onto the plane $m_1$--$m_{ee}$. The vertical grey bands
indicate the excluded regions from the absolute neutrino mass scale upper bound from 
cosmology, $0\nu\b\b$ and Tritium beta decay given in the text. The horizontal grey band
is the excluded region from the conservative upper bound on $m_{ee}$, while the horizontal dot-dashed line is the 
most stringent one (see Eq.~(\ref{upperbmee})). The solid lines denote the usual full accessible region.}
\end{figure}
It can be noticed how, even in the case of NO, there is a lower bound on $m_{ee}\gtrsim 0.1\,{\rm meV}$.
The bulk of points ($\sim 95\%$) fall however in the range (1--30) meV that will be significantly tested in future
years by new generation $0\nu\beta\beta$ experiments. 
\begin{figure}[t]
\begin{center}
\psfig{file=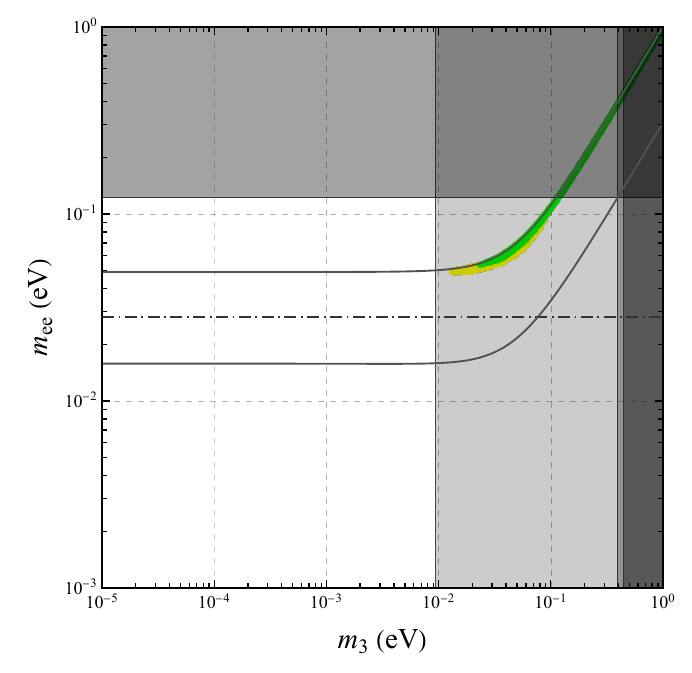,scale=0.5} \hspace{-0mm}
\psfig{file=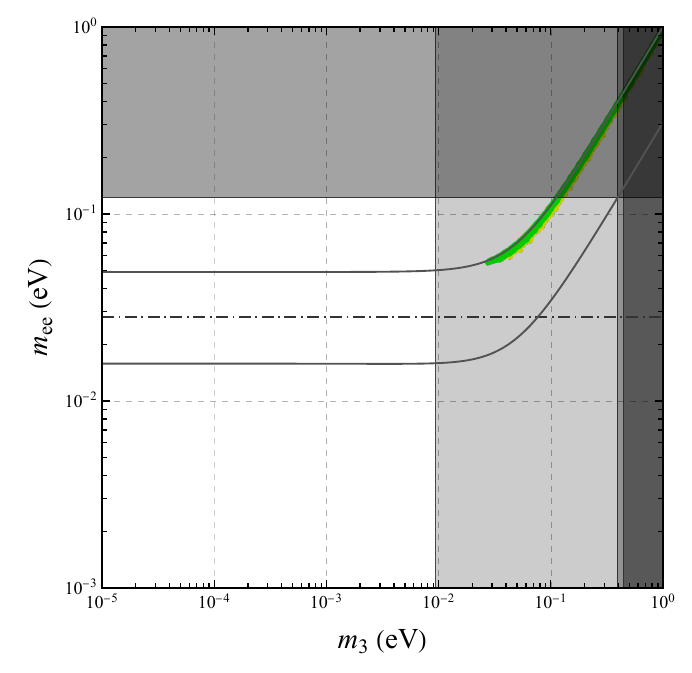,scale=0.5} 
\end{center}
\vspace{-8mm}
\caption{Projection of the scatter plot for IO onto the plane $m_1$--$m_{ee}$. Same conventions as in previous figure.}
\end{figure}
Finally, in the left panels of Fig.~9 and Fig.~10, for NO and IO respectively, we show the corresponding scatter plots for the
heavy neutrino masses.
\begin{figure}[t]
\begin{center}
\psfig{file=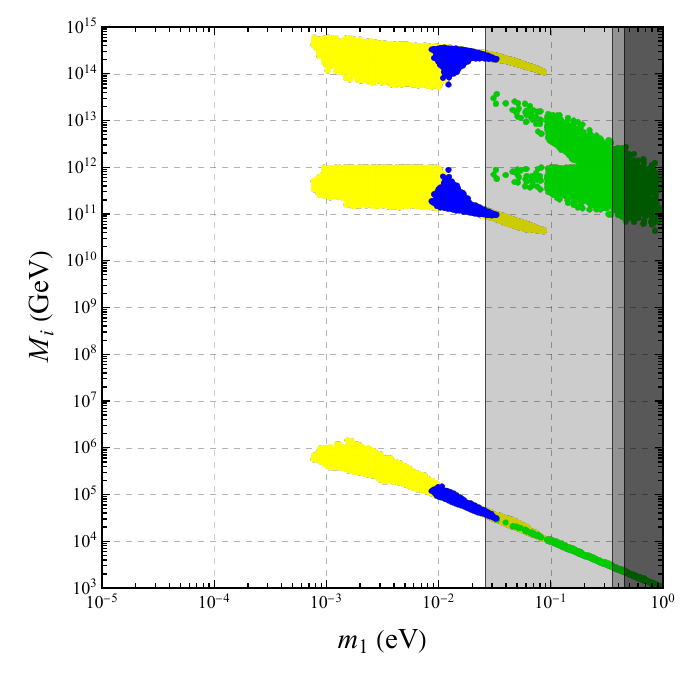,scale=0.5} \hspace{-0mm}
\psfig{file=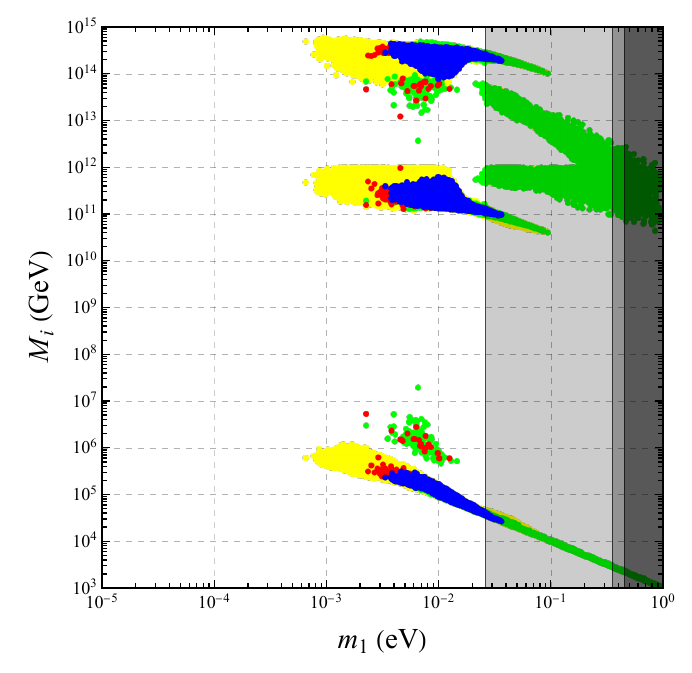,scale=0.5} 
\end{center}
\vspace{-8mm}
\caption{Scatter plots, corresponding to those in Fig.~3 for NO, for the seesaw neutrino mass spectrum.}
\end{figure}
In the case of NO, the solutions should be compared to the points in the top right panel In Fig.~1 obtained without imposing the successful
leptogenesis condition. The most clear effect of imposing successful $SO(10)$-inspired leptogenesis is, once more, the 
lower bound on $m_1 \gtrsim 1\,{\rm meV}$. The other feature is that the allowed bands for the $M_i$'s 
considerably narrow. The main reason for that is that the two Majorana phases and Dirac phase now get correlated and
their allowed regions are around special values. For example, if one looks at the expression for $M_1$ in 
Eq.~(\ref{RHspectrum}), and taking approximately $V_L \simeq I$ so that $(\widetilde{m}_\nu)_{11} \simeq m_{ee}$, 
this narrow band for $M_1$ clearly stems for the much more restricted values of $m_{ee}$ versus $m_1$. This is well understood
analytically since one needs $K_{1\tau} (K_{1\mu}) \lesssim 1$ for tauonic (muonic) solutions and this implies correlations among phases.  
\begin{figure}[t]
\begin{center}
\psfig{file=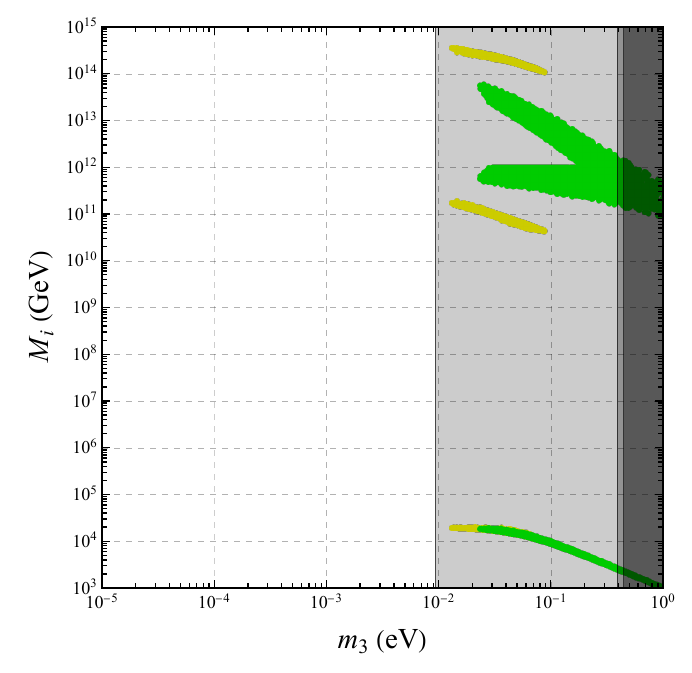,scale=0.5} \hspace{-0mm}
\psfig{file=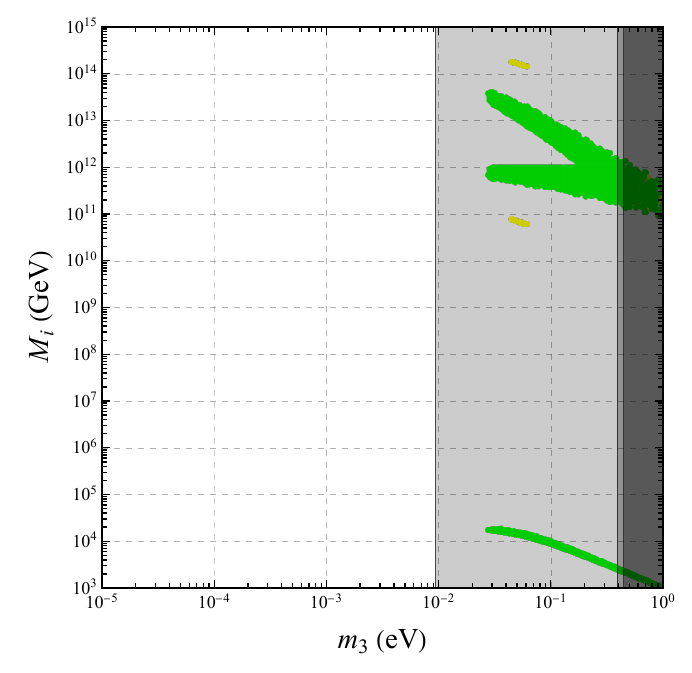,scale=0.5} 
\end{center}
\vspace{-8mm}
\caption{As in previous figure but for IO.}
\end{figure}

\subsubsection{Accounting for flavour coupling}

We can now also take into account flavour coupling originating from spectator processes \cite{Barbieri:1999ma,Buchmuller:2001sr}. 
The expression for the final $B-L$ asymmetry now contains 27 terms, 9 terms for each flavoured asymmetry ($\a =e,\mu,\tau$) 
\cite{fuller,DiBari:2025zlv}:
\bea\label{NfDa2}
N^{\rm f}_{\D_{\a}}
 & = &  V^{-1}_{\a e''}\,
\left[\sum_{\b}\,V_{e''\b}\,N_{\D_{\b}}^{T\sim T_L}\right]
\,e^{-{3\pi\over 8}\,K_{1 e''}} \\  \nonumber
& + &  V^{-1}_{\a \m''}\,
\left[\sum_{\b}\,V_{\m''\b}\,N_{\D_{\b}}^{T\sim T_L}\right]
\,e^{-{3\pi\over 8}\,K_{1 \m''}} \\ \nonumber
& + &  V^{-1}_{\a \t''}\,
\left[\sum_{\b}\,V_{\t''\b}\,N_{\D_{\b}}^{T\sim T_L}\right]
\,e^{-{3\pi\over 8}\,K_{1 \t''}} \,  ,
\eea
and $N^{\rm f}_{B-L} = N^{\rm f}_{\D_{e}} + N^{\rm f}_{\D_{\mu}} + N^{\rm f}_{\D_{\tau}}$.
The quantities $N_{\D_{\b}}^{T\sim T_L} \, (\beta =e,\mu,\tau)$ are the flavoured asymmetries at the production, while
the $K_{1 \m''}$ are the rotated flavoured decay parameters. We refer to \cite{DiBari:2025zlv} for 
the explicit expressions and definitions of these quantities and 
for the matrix elements $V_{\alpha'',\beta}$ and $V^{-1}_{\a \beta''}$ operating a rotation 
from the flavoured basis to the rotated flavoured basis and its inverse.
In the right panels of Fig.~3 and Fig.~4, for NO and IO respectively, we show again the scatter plots projected 
in the three-dimensional $(m_1,\delta,\theta_{23})$ space. These are
obtained following the same procedure described for the left panels but this time using Eqs.~(\ref{NfDa2}) for the calculation of the final $B-L$
asymmetry accounting for flavour coupling. The results confirm again those obtained in \cite{DiBari:2025zlv}. 
In particular, the main effect of flavour coupling is the appearance of new muonic solutions (and even some very marginal electronic
solutions). One can also notice how flavour coupling slightly relaxes the lower bound on $m_1$, from $0.89\,{\rm meV}$ to $0.65\,{\rm meV}$.

\subsection{Strong thermal $SO(10)$-inspired leptogenesis}

We have so far assumed that leptogenesis is the only mechanism responsible for a $B-L$ asymmetry production. 
However, more generally, the  final asymmetry is given by the sum of two terms,
\be\label{2terms}
N_{B-L}^{\rm f} = N_{B-L}^{\rm p,f} + N_{\rm B-L}^{\rm lep,f}  \,  .
\ee
The first term is the relic value of a possible pre-exisisting asymmetry generated prior to the onset of leptogenesis by some external mechanism. 
The second is the contribution from leptogenesis. Correspondingly, the baryon-to-photon number ratio is also given, in general,  
by the sum of two contributions, $\eta_B^{\rm p}$ and $\eta_B^{\rm lep}$, respectively. 
We can denote by $N_{B-L}^{\rm p,i}$ the initial value of the pre-existing $B-L$ asymmetry, that could be regarded as constant 
between the end of inflation and the onset of leptogenesis.
In the absence of any washout, this would translate into 
a unphysical sizeable value of $\eta_B^{\rm p}$ comparable or greater than $\eta_B^{\rm exp}$.
A strong thermal leptogenesis condition requires that the initial pre-existing
asymmetry is efficiently washed out by seesaw neutrino inverse processes,
in a way that $N_{B-L}^{\rm p,f} \ll N_{\rm B-L}^{\rm lep,f}$. 

the relic value of the pre-existing asymmetry can also be expressed as the sum of three contributions from each flavour 
i.e., $N_{B-L}^{\rm p, f} = \sum_{\a} \,  N_{\D_\a}^{\rm p,f}$.
Neglecting flavour coupling, the expressions for the $N_{\D_\a}^{\rm p,f}$ are given by \cite{STSO10,chianese}
\bea\label{finalpas}
N_{\D_\t}^{\rm p,f} & = & 
(p^0_{{\rm p}\t}+\D p_{{\rm p}\tau})\,  e^{-{3\pi\over 8}\,(K_{1\t}+K_{2\t})} \, N_{B-L}^{\rm p,i} \,  , \\  \nonumber
N_{\D_\m}^{\rm p,f} & = & 
\left\{(1-p^0_{{\rm p}\t})\,\left[
p^0_{\mu\t_2^{\bot}}\, p^0_{{\rm p}\t^\bot_2}\,
e^{-{3\pi\over 8}\,(K_{2e}+K_{2\m})} + (1-p^0_{\m\t_2^{\bot}})\,(1-p^0_{{\rm p}\t^\bot_2}) \right]  + 
\D p_{{\rm p}\mu}\right\}
\,e^{-{3\pi\over 8}\,K_{1\m}}\, N_{B-L}^{\rm p,i}
 ,  \\  \nonumber
N_{\D_e}^{\rm p,f}& = & 
\left\{(1-p^0_{{\rm p}\t})\,\left[ 
p^0_{e\t_2^{\bot}}\,p^0_{{\rm p}\t^\bot_2}\,
e^{-{3\pi\over 8}\,(K_{2e}+K_{2\m})} + (1-p^0_{e \t_2^{\bot}})\,(1-p^0_{{\rm p}\t^\bot_2}) \right]  + \D p_{{\rm p} e}\right\}
 \,e^{-{3\pi\over 8}\,K_{1e}} \,\, N_{B-L}^{\rm p,i} \,   .
\eea
The quantities $p^0_{{\rm p} \tau}$ and $p^0_{{\rm p}\tau_2^{\bot}}$ are
the fractions of the initial pre-existing asymmetry in the tauon flavour and 
in the flavour $\tau_2^{\bot}$, where $\tau_2^{\bot}$  is the  electron and muon flavour  
superposition component in the leptons  
produced by $N_2$-decays (or equivalently the flavour component that is 
washed-out in the $N_2$-inverse processes).\footnote{Notice that in general $p^0_{{\rm p} \tau}+
p^0_{{\rm p}\tau_2^{\bot}} \neq 1$. This is because in general the projection of the pre-existing asymmetry flavour direction on the
$\tau^\bot$ plane, is not parallel to $\tau^\bot_2$.}
We also introduced the two quantities $p^0_{\alpha \t_2^{\bot}} \equiv K_{2\alpha}/(K_{2e}+K_{2\mu}) \; (\a=e,\mu)$. These give
the fractions of the $\a$ flavour  in the  $\tau_2^{\bot}$ component of the lepton quantum state $|\ell_2\rangle$ produced by $N_2$-decays.
The terms $\D p_{{\rm p} \t},\D p_{{\rm p} e}$ and $\D p_{{\rm p}\mu}$ are so-called phantom terms and 
such that $\D p_{{\rm p} e}+\D p_{{\rm p}\mu} +  \D p_{{\rm p}\tau} = 0$. 
They take into account the possibility that the mechanism
that created the pre-existing asymmetry produced lepton states that  are not exactly the $C\!P$ conjugated of the anti-lepton states. 
It can be crossed-checked that in Eq.~(\ref{finalpas}), if one switches-off the washout terms, 
the sum of the three final flavoured pre-existing asymmetries
is equal to the total initial pre-existing asymmetry, as it should be.

Satisfying simultaneously strong thermal and successful leptogenesis conditions is highly non-trivial \cite{problem}.
For this reason, it is quite remarkable that there exists a subset of the solutions satisfying successful $SO(10)$-inspired 
leptogenesis that simultaneously satisfies also the strong thermal leptogenesis condition \cite{STSO10}. 
We have determined such subset in the case $N_{B-L}^{\rm lep,i} = 10^{-3}$
These solutions are indicated  with blue circles in all scatter plots. The most striking feature is that for this subset of solutions the
lower bound on $m_1$ and $m_{ee}$ becomes much more stringent and more easily testable.
Neglecting flavour coupling one has $m_1, m_{ee} \gtrsim 10\,{\rm meV}$.

Something that
can be noticed very clearly is that such subset is absent in the IO case. The reason
why the strong thermal condition cannot be realised for IO within $SO(10)$-inspired leptogenesis
has been clearly explained in \cite{decrypting}, where it was shown that the muonic component 
of the pre-existing asymmetry necessarily escapes the washout.

\section{Extended $SO(10)$-inspired leptogenesis}

Let us now finally consider the case when the matrix $V_L$ is left completely free. 
It is quite remarkable that the analytical expressions for the seesaw neutrino masses $M_I$,
the RH neutrino mixing matrix $U_R$ and even for the orthogonal matrix $\Omega$ are still valid (still barring of course
vicinity of crossing level solutions).  
In the two panels of Fig.~11, neglecting (left panel) or including (right panel) flavour coupling,
we show for NO a scatter plot that is again a random scan performed following the same prescriptions
listed in the previous section except that this time we scan on the three angles in $V_L$ in the entire interval $[0,\pi]$.
\begin{figure}[t]
\begin{center}
\psfig{file=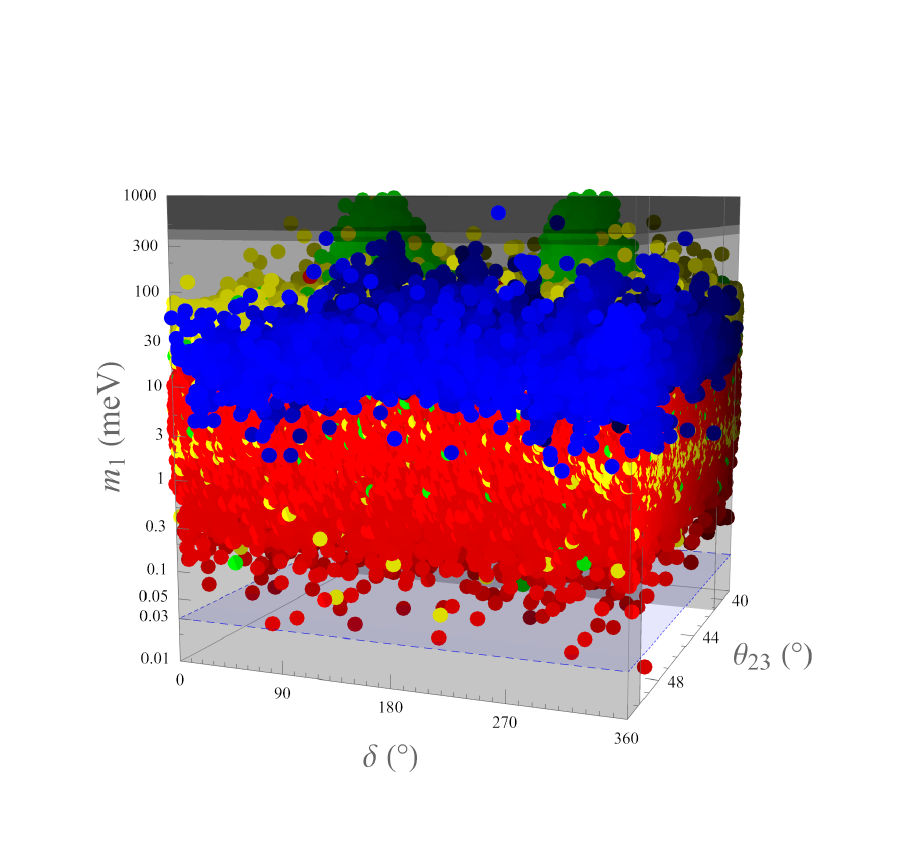,scale=0.49} \hspace{-10mm}
\psfig{file=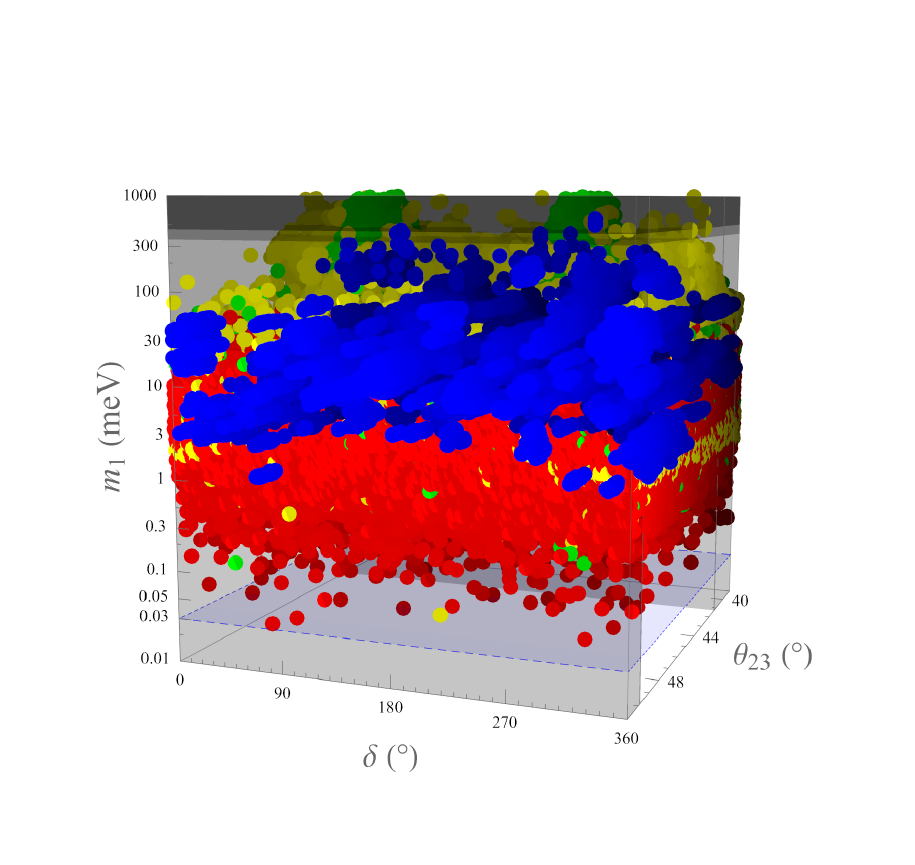,scale=0.49}
\end{center}
\vspace{-10mm}
\caption{Projection on the three-dimensional space $(m_1,\delta,\theta_{23})$ of the
scatter plots of the solutions obtained for arbitrary $V_L$ in the case of NO, imposing successful $SO(10)$-inspired leptogenesis,
neglecting (left) and including (right) flavour coupling. Conventions are the same as in Fig.~3.}
\end{figure}
One can see now how the full region is filled but there is still a lower bound $m_1 \gtrsim 0.03\,{\rm meV}$.
The strong thermal leptogenesis subset of solutions also considerably enlarges.
\begin{figure}[t]
\begin{center}
\vspace{-8mm}
\psfig{file=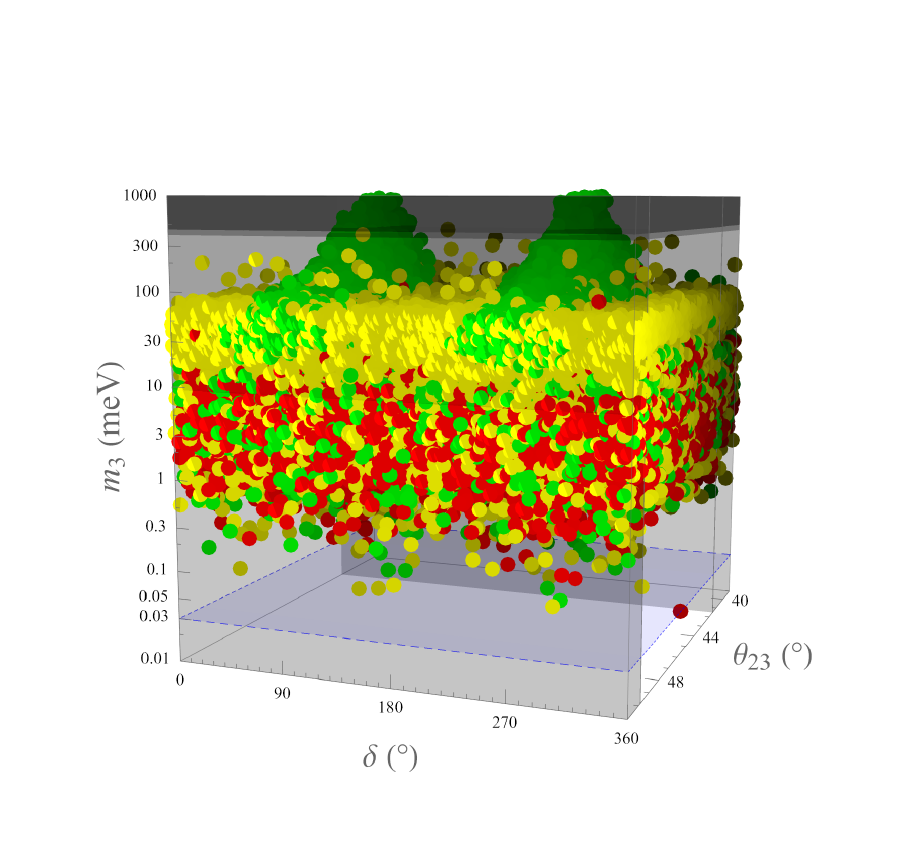,scale=0.5} \hspace{-10mm}
\psfig{file=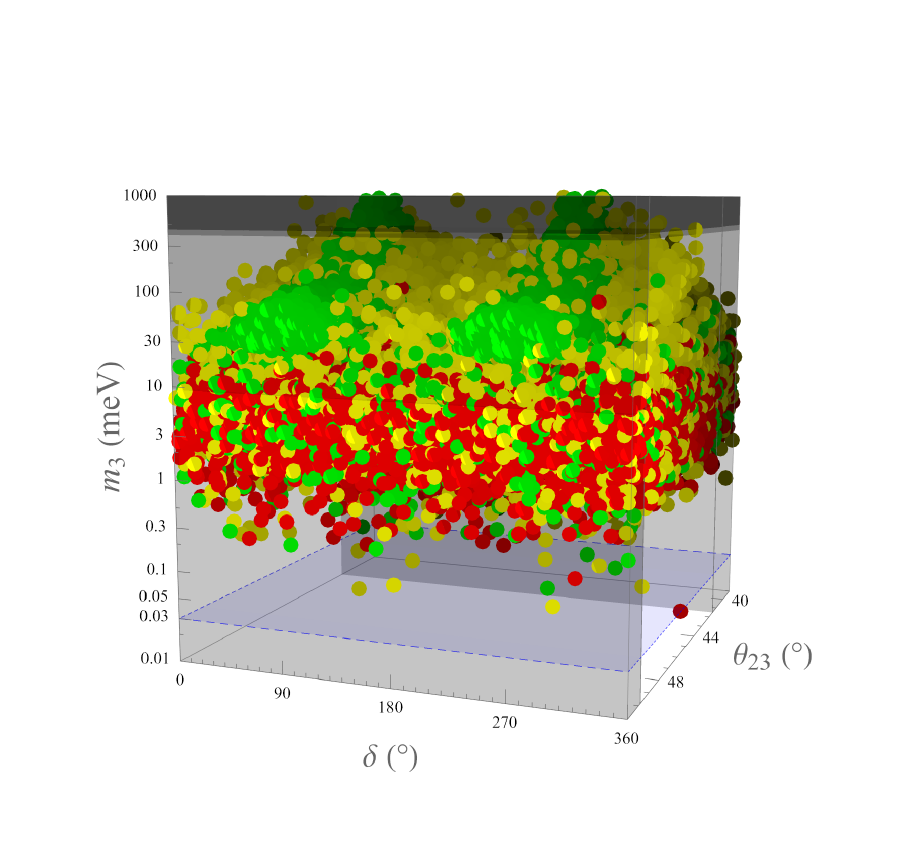,scale=0.5}
\end{center}
\vspace{-10mm}
\caption{As in previous figure but for IO.}
\end{figure}
In Fig.~12, neglecting (left) and including (right) flavour coupling, we show the results for IO. 
It is clear how now there are many viable solutions as for NO, the lower bound  we find is practically the same.
However, the most important difference, is that there are no strong thermal solutions (no blue points) for IO.
In this way we can conclude that even for extended $SO(10)$-inspired leptogenesis, the strong thermal leptogenesis
condition necessarily requires NO. This is an interesting result especially in the light of the current
neutrino oscillation data that start to strongly favour NO. 
 
Another  difference is that while in the case of NO there is somehow a balance in the flavour distribution, 
with some preponderance of electronic solutions, in the IO case electronic solutions are in fact quite marginal. 
All these features are even more evident from the following plots, where we
continue to show results both neglecting (left panels) and including (right panels) flavour coupling. 
\begin{figure}[t]
\begin{center}
\psfig{file=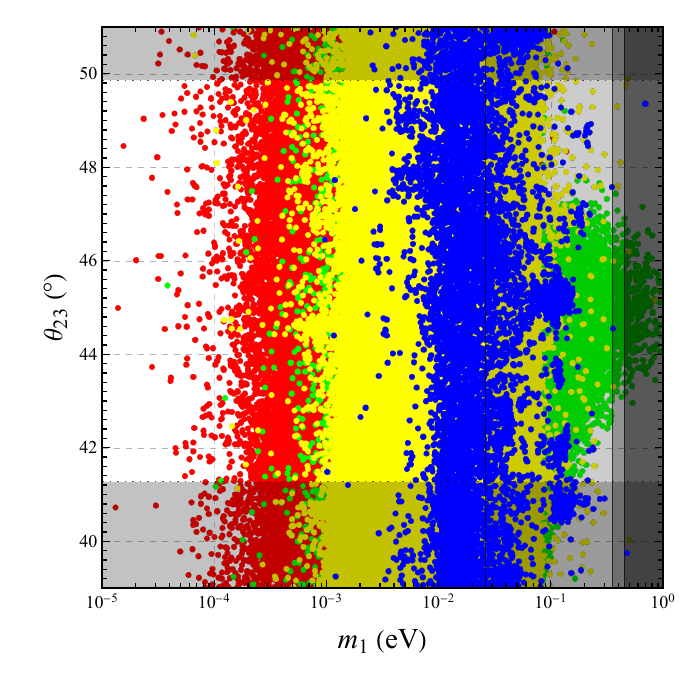,scale=0.5} \hspace{-0mm}
\psfig{file=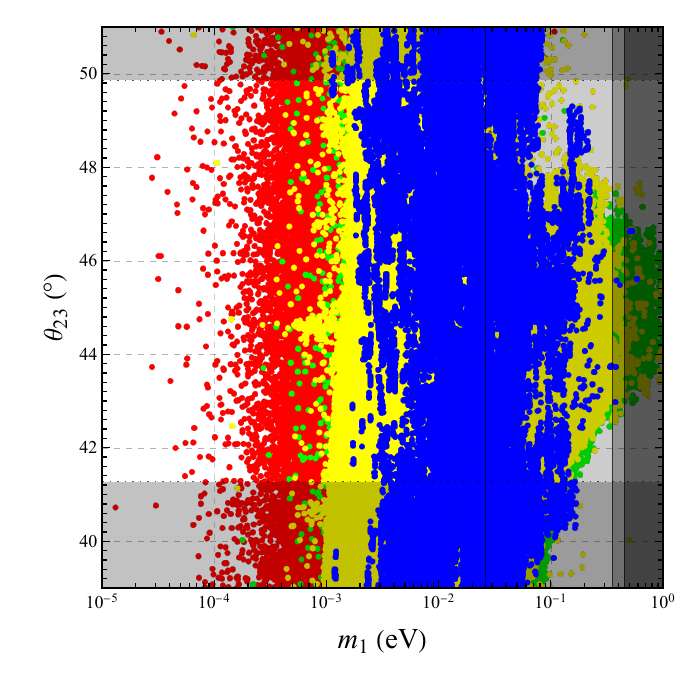,scale=0.5}
\end{center}
\vspace{-8mm}
\caption{Projection of the scatter plot for arbitrary $V_L$ on the plane $m_1$--$\theta_{23}$ for NO.}
\end{figure}
In Fig.~13 and Fig.~14 we show again the projection of the scatter plot in the plane $\theta_{23}$ versus $m_1$
for NO and IO, respectively. It is clear how now there are no excluded values of $\theta_{23}$ for $m_1 \gtrsim 1\,{\rm meV}$.
\begin{figure}[t]
\begin{center}
\psfig{file=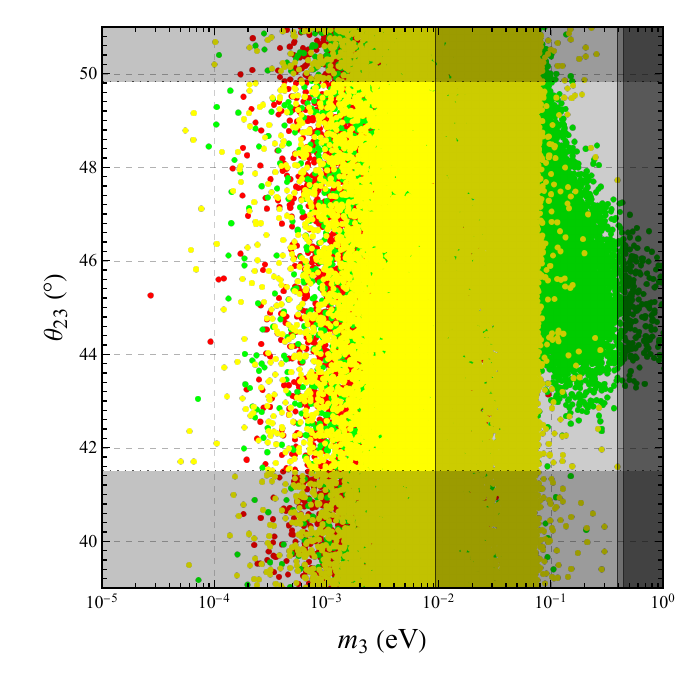,scale=0.5} \hspace{-0mm}
\psfig{file=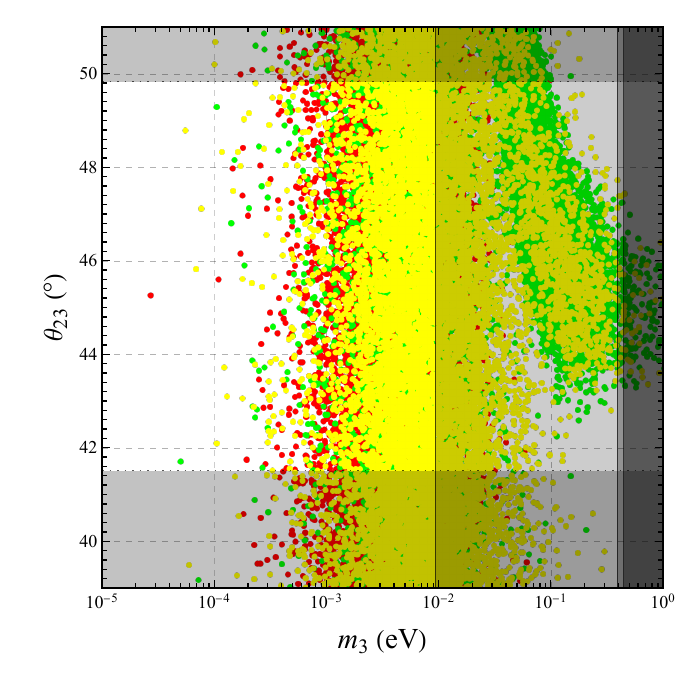,scale=0.5}
\end{center}
\vspace{-8mm}
\caption{As in previous figure but for IO.}
\end{figure}
One can clearly notice how electronic solutions are very marginal for IO.  

In Fig. 15 and Fig. 16, for NO and IO respectively, we show the projections of the scatter plot onto the important
plane $m_{ee}$ versus $m_1$.  For NO there is a significant relaxation of the lower bound on $m_{ee}$, as on $m_1$.
However, also in this case there is still a similar lower bound $m_{ee} \gtrsim 10^{-5}\,{\rm eV}$.
\begin{figure}[t]
\begin{center}
\vspace{-0mm}
\psfig{file=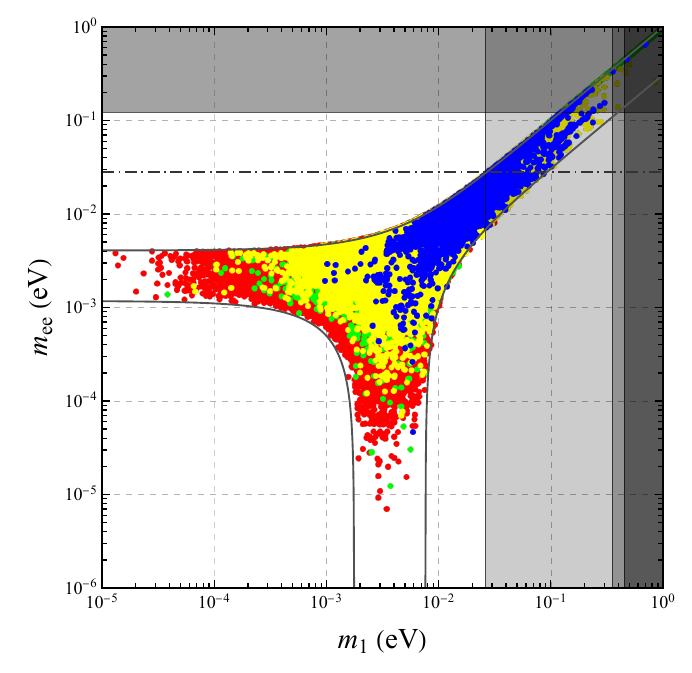,scale=0.5} \hspace{-0mm}
\psfig{file=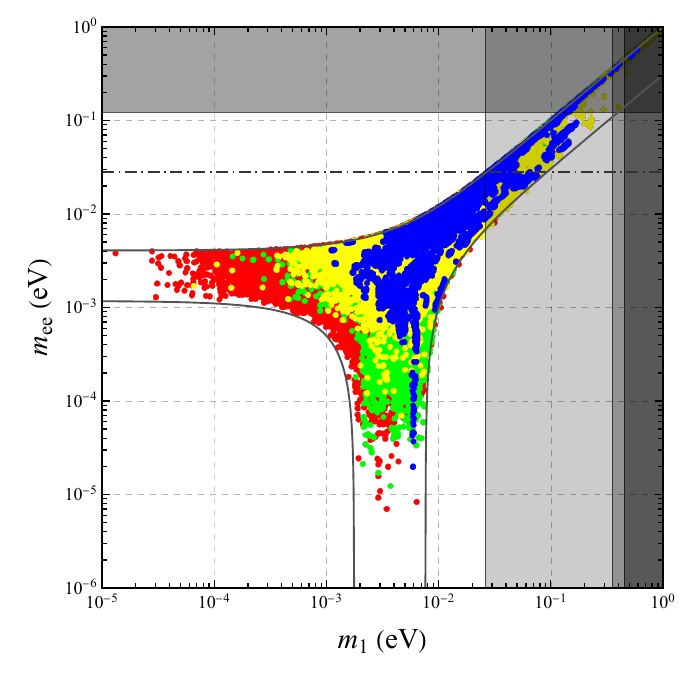,scale=0.5}
\end{center}
\vspace{-8mm}
\caption{Projection of the scatter plot for arbitrary $V_L$ on the plane $m_1$--$m_{ee}$ for NO.}
\end{figure}
For IO the solutions fill approximately all the physically allowed region. 
\begin{figure}[t]
\begin{center}
\vspace{-0mm}
\psfig{file=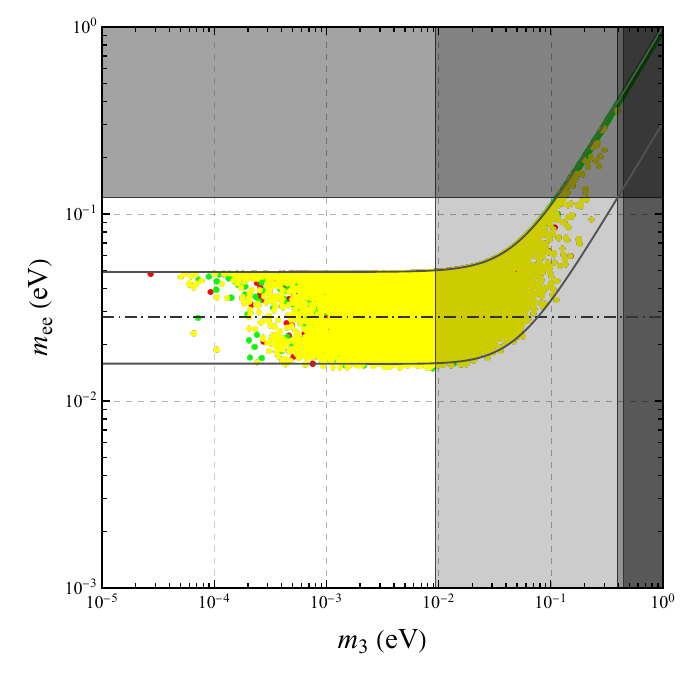,scale=0.5} \hspace{-0mm}
\psfig{file=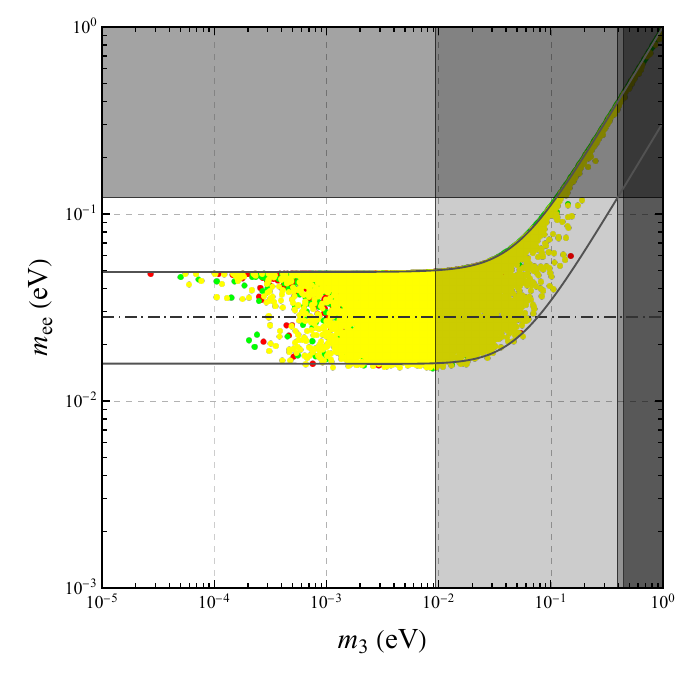,scale=0.5}
\end{center}
\vspace{-8mm}
\caption{As in previous figure but for IO.}
\end{figure}
Finally in Fig.~17 and Fig.~18, for NO and IO respectively, we show the scatter plots for the seesaw neutrino masses
in the extended random scan of parameters. Both in NO and IO, if one compares with the botton left panels in Fig.~1 and Fig.~2,
it is quite evident how there are no solutions for $M_2 \lesssim 10^{11}\,{\rm GeV}$. This is quite clear, since one
needs a sufficiently heavy $N_2$ neutrino to get sufficiently large $C\!P$ asymmetries to satisfy successful 
leptogenesis condition.  
\begin{figure}[t]
\begin{center}
\vspace{-0mm}
\psfig{file=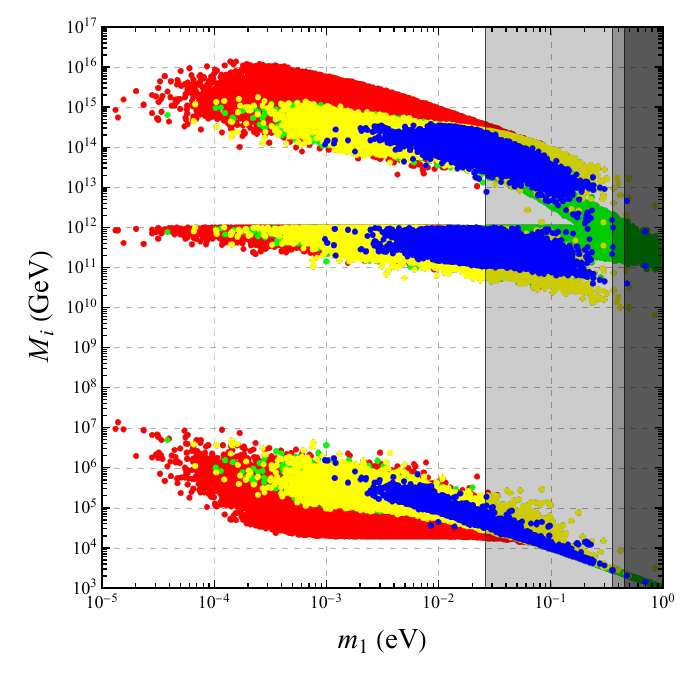,scale=0.5} \hspace{-0mm}
\psfig{file=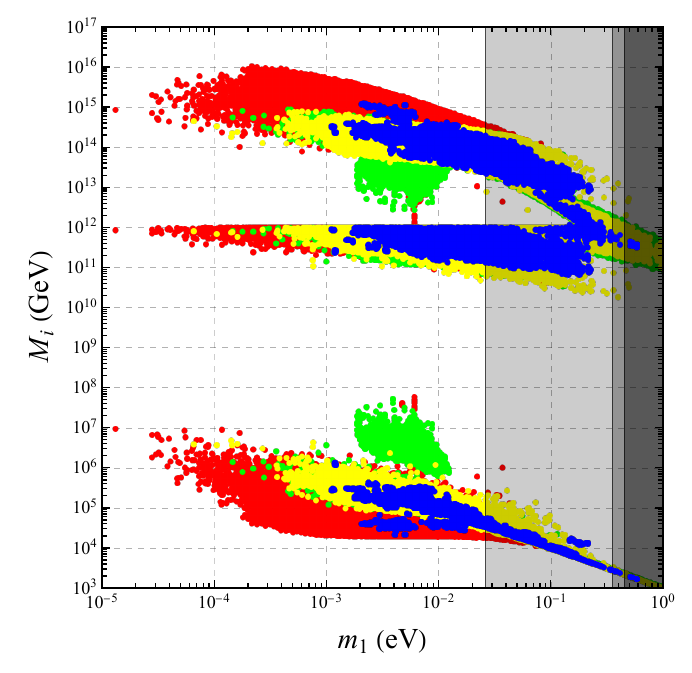,scale=0.5}
\end{center}
\vspace{-8mm}
\caption{Scatter plot of heavy neutrino masses for arbitrary $V_L$ and NO.}
\end{figure}
This also explains the total absence of solutions for $M_1 \gtrsim 10^7\,{\rm GeV}$, the two 
features are closely linked. This is clear form the analytical expression for $M_1$ and $M_2$ in Eq.~(\ref{RHspectrum}):
large $M_1$ values are obtained for low values of $|(\widetilde{m}_\nu)_{11}|$ and these in turn would necessarily imply
low values of $M_2$. 
\begin{figure}[t]
\begin{center}
\vspace{-0mm}
\psfig{file=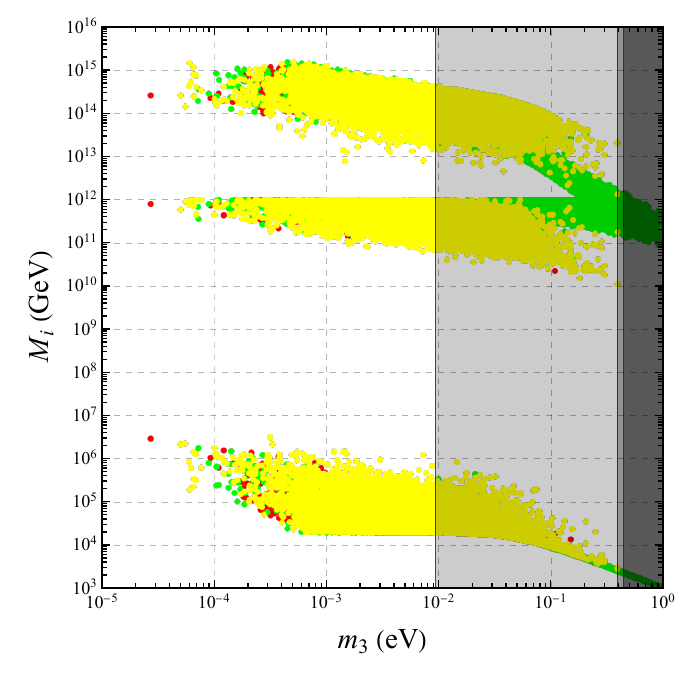,scale=0.5} \hspace{-0mm}
\psfig{file=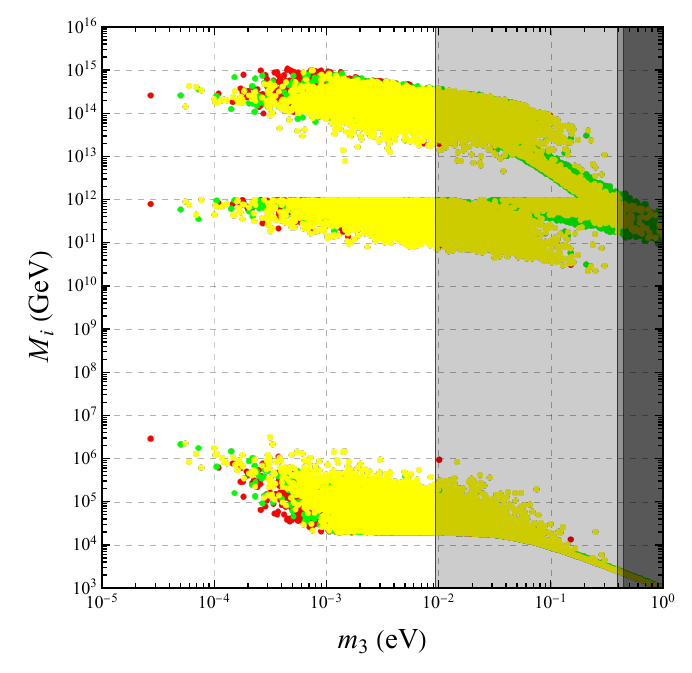,scale=0.5}
\end{center}
\vspace{-8mm}
\caption{Scatter plot of heavy neutrino masses for arbitrary $V_L$ and IO.}
\end{figure}
It is not excluded that solutions close to crossing level solution for $M_1 \sim M_2$, though extremely fine-tuned,
might appear. This would require to describe the generation of the asymmetry in the quasi-degenerate case, taking into account
combined $N_1$ and $N_2$ production and wash-out.\footnote{Recently, such a kind of solution has been found in \cite{He:2026xdy}
for Dirac neutrino masses equal to down quark masses. The requested quasi-degeneracy is $(M_2 - M_1)/M_2 \sim 10^{-5}$.}

\subsection{Turning on just the $\theta_{23}^{\rm L}$ angle in $V_L$: rise and fall of muonic solutions}

Here we want to give some insight in the results we have shown. Having turned on simultaneously all angles in $V_L$,
letting them free to vary, has clearly opened a whelm of new solutions. Here we would like to understand better the reason why. 
We focus on NO for definiteness and since this is much more attractive and favoured by the data. 
In Fig.~19 we show scatter plots of solutions satisfying successful leptogenesis for 
a $V_L$ such that $\theta_{12}^{\rm L} < \theta_{12}^{\rm CKM}$ and $\theta_{13}^{\rm L} < \theta_{13}^{\rm CKM}$ while
the values of $\theta_{23}^{\rm L}$ vary in the panels as described in the figure caption.
\begin{figure}[t]
\begin{center}
\psfig{file=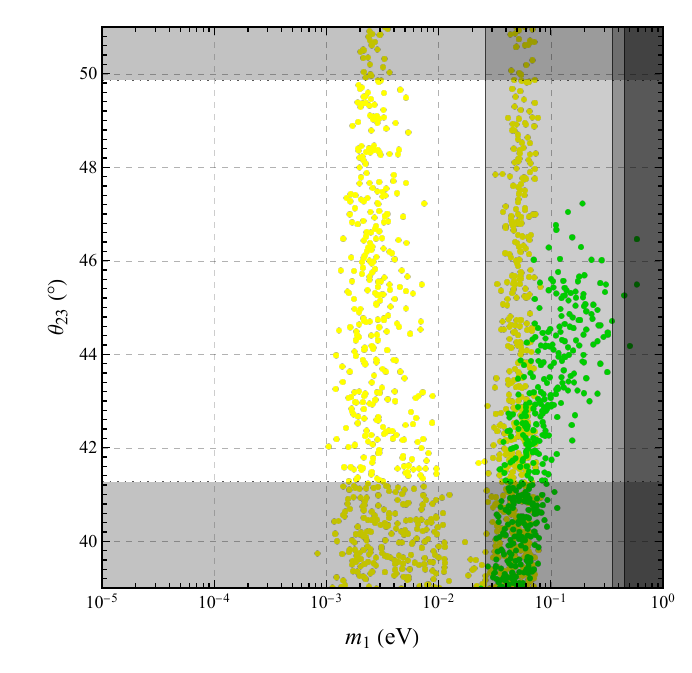,scale=0.2}
\psfig{file=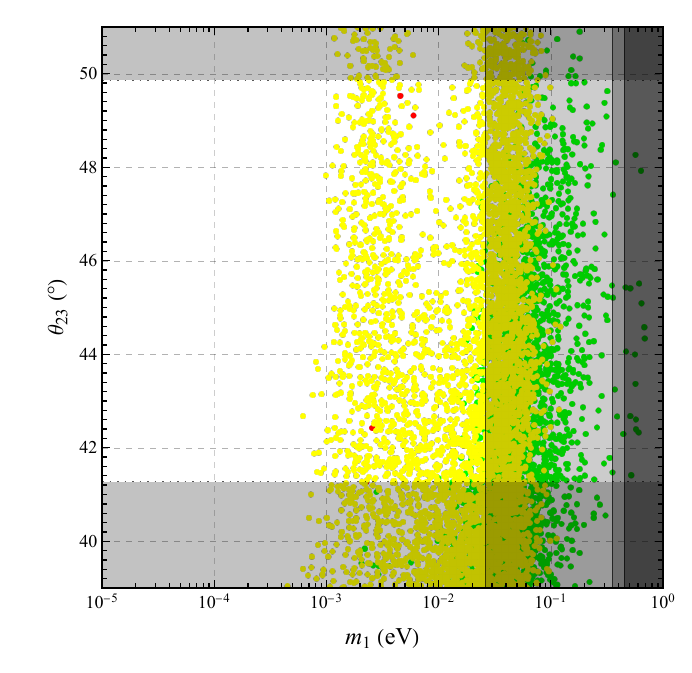,scale=0.2} 
\psfig{file=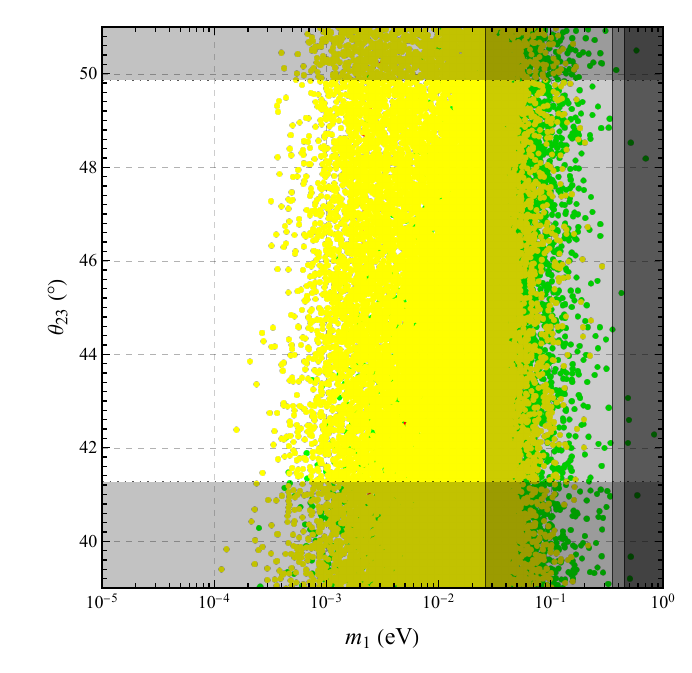,scale=0.2} 
\psfig{file=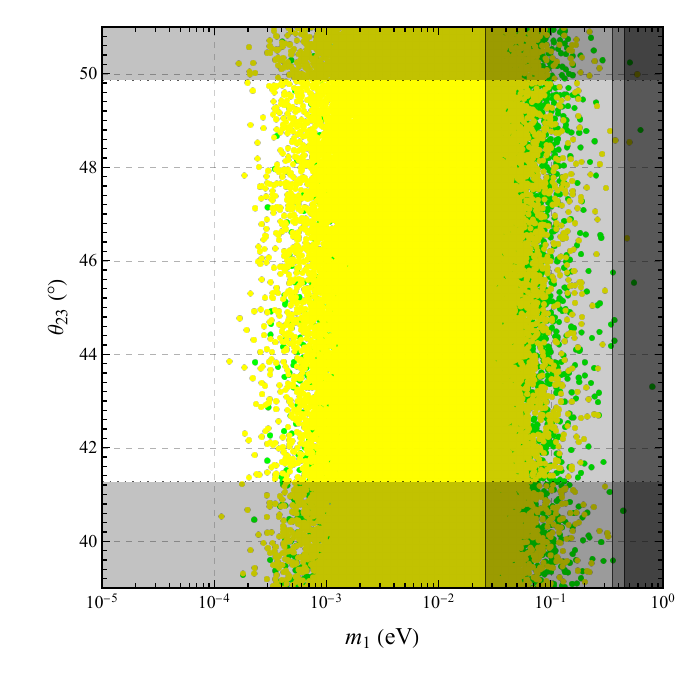,scale=0.2} \\ 
\psfig{file=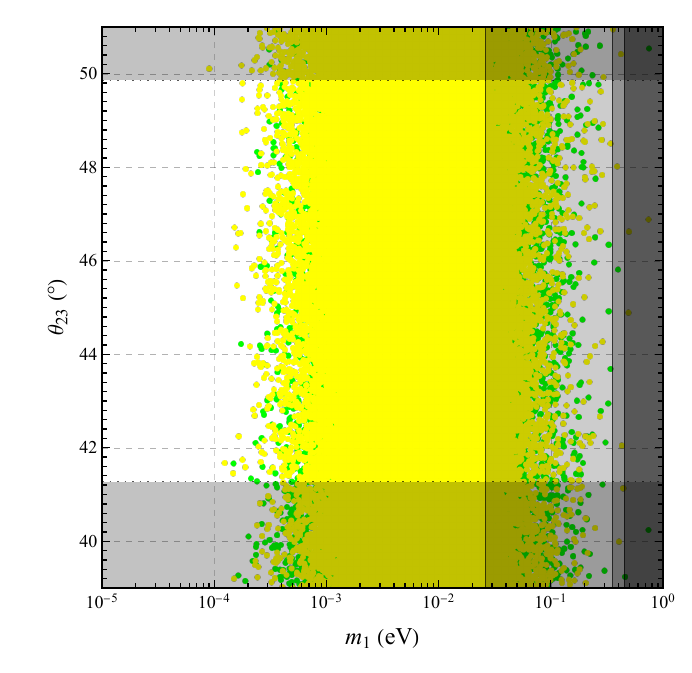,scale=0.2} 
\psfig{file=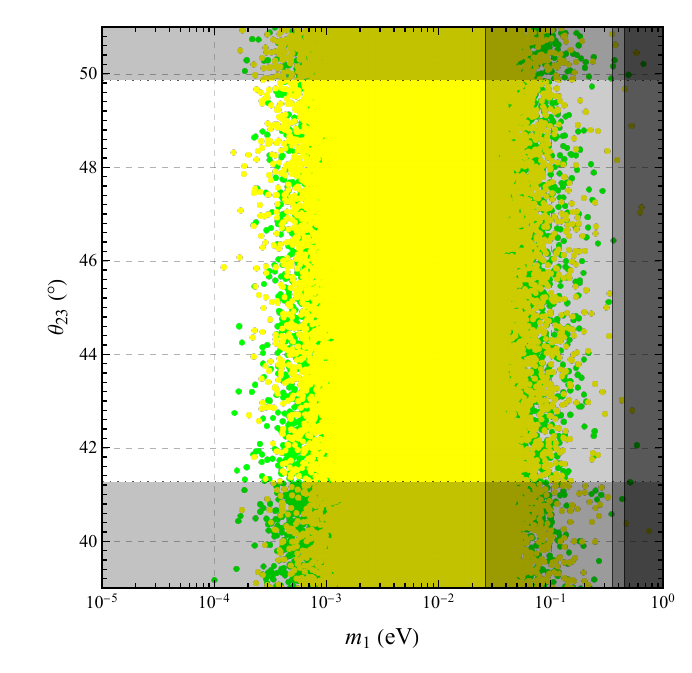,scale=0.2} 
\psfig{file=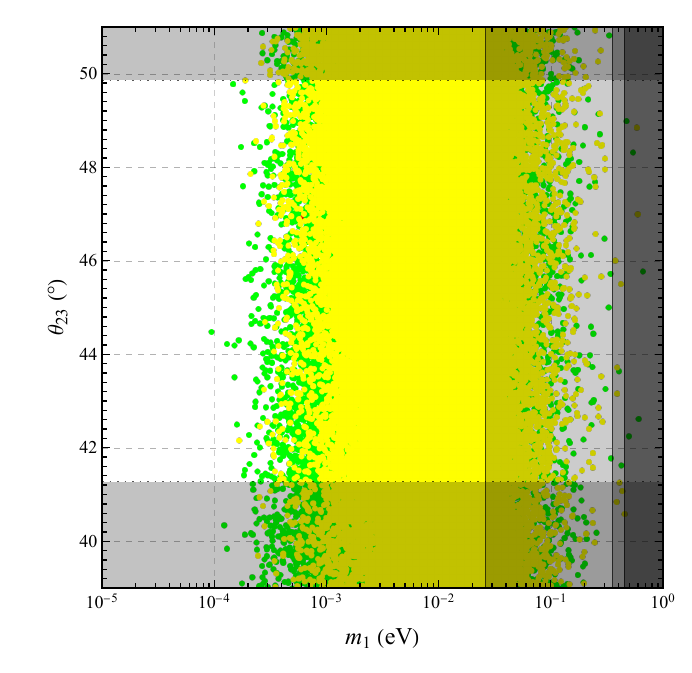,scale=0.2} 
\psfig{file=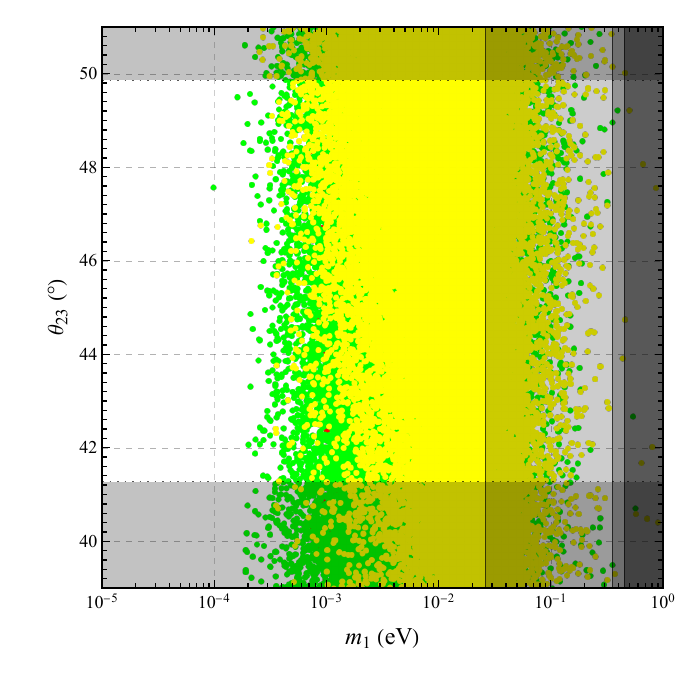,scale=0.2} \\ 
\psfig{file=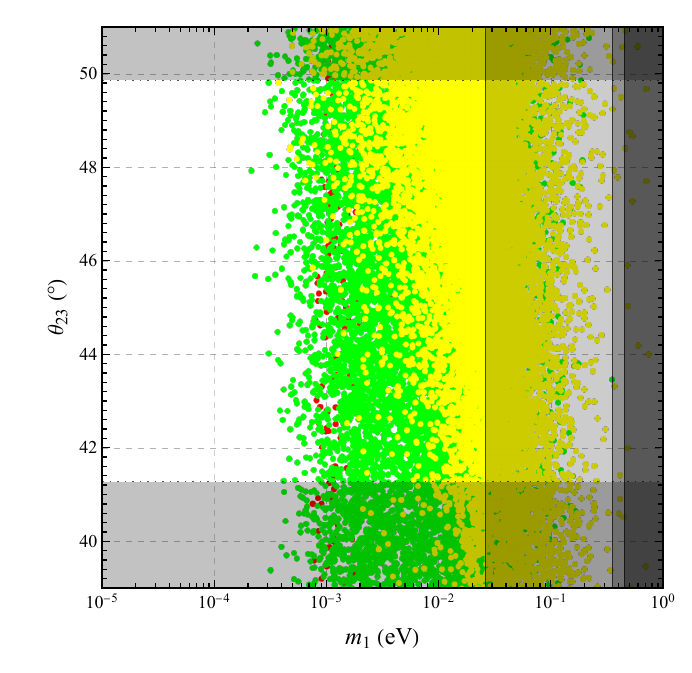,scale=0.2} 
\psfig{file=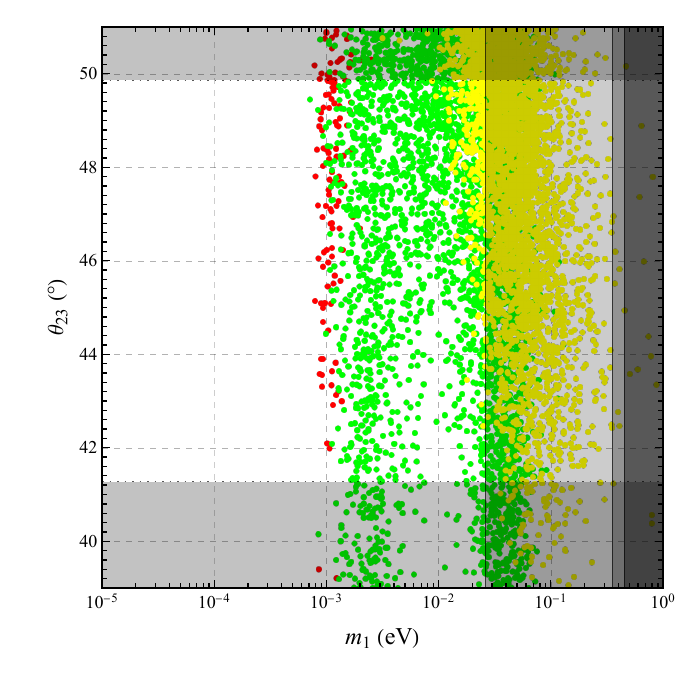,scale=0.2} 
\psfig{file=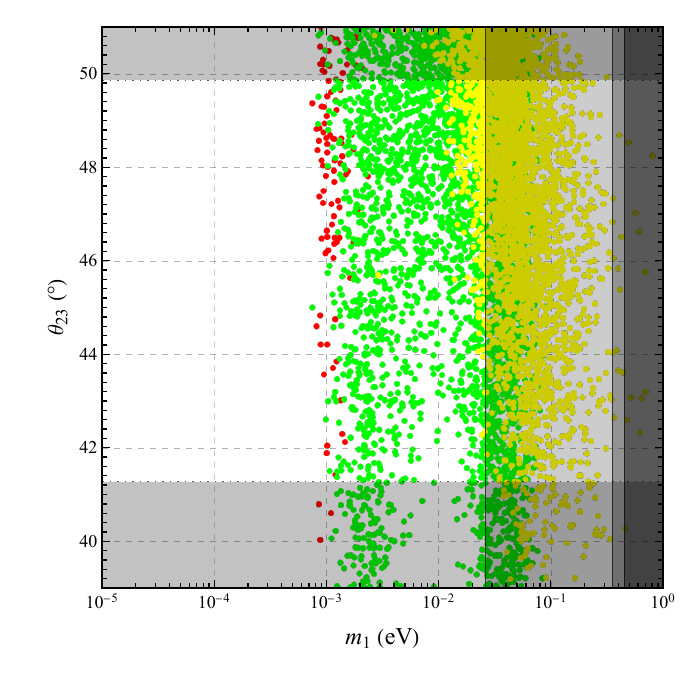,scale=0.2} 
\psfig{file=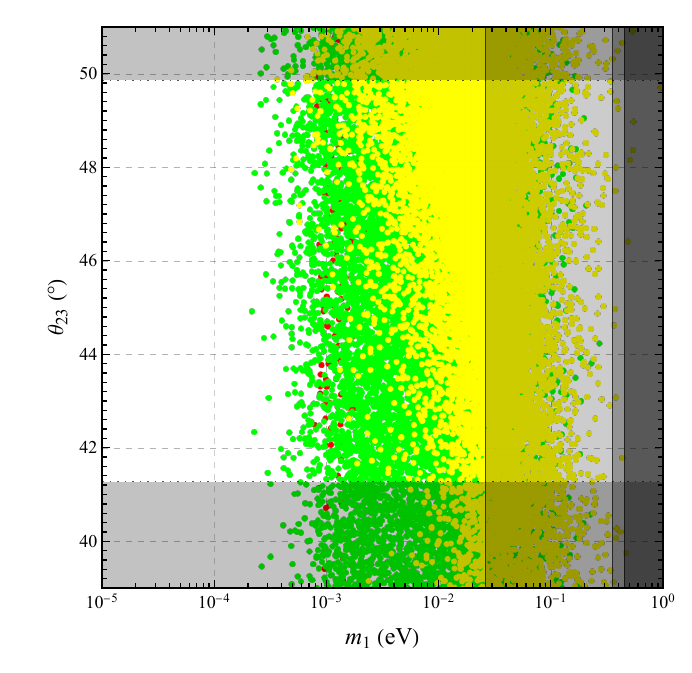,scale=0.2} \\ 
\psfig{file=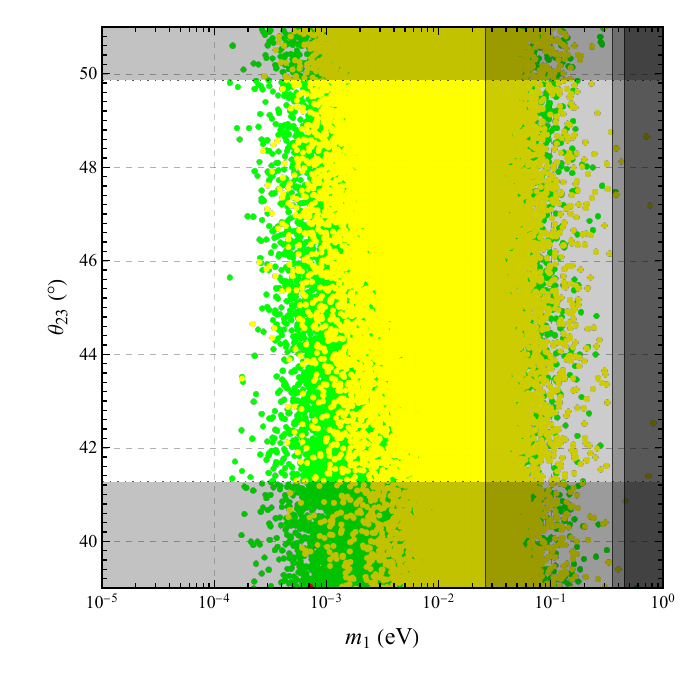,scale=0.2} 
\psfig{file=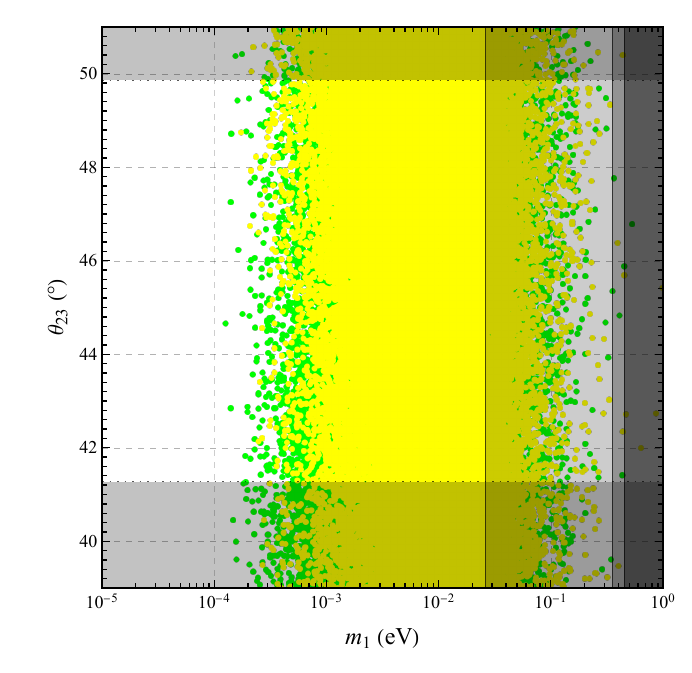,scale=0.2} 
\psfig{file=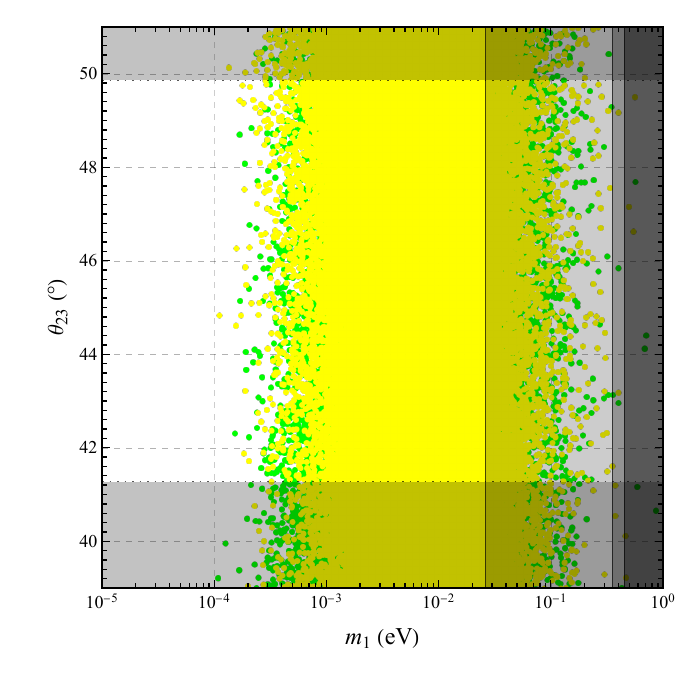,scale=0.2} 
\psfig{file=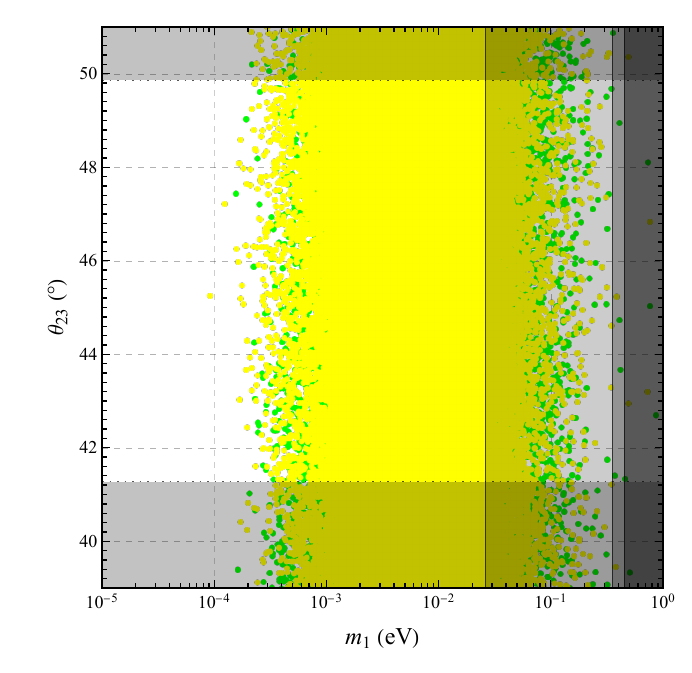,scale=0.2} \\ 
\psfig{file=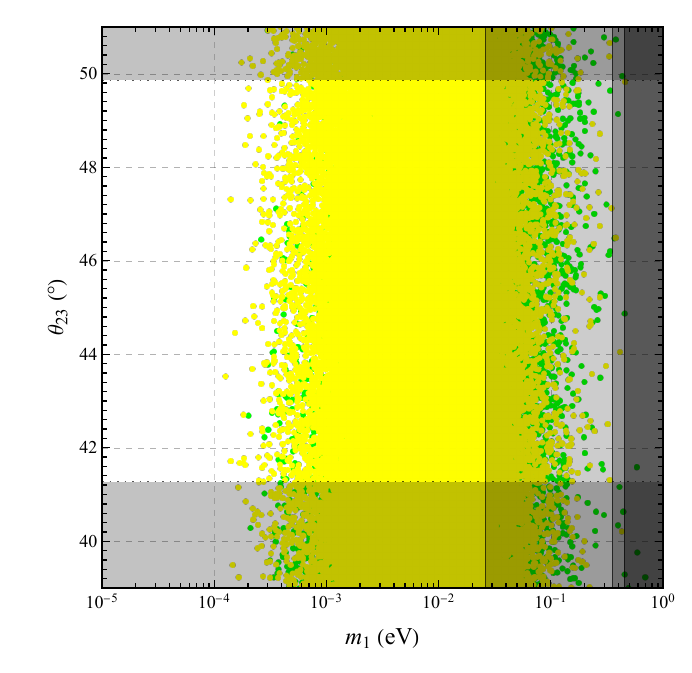,scale=0.2} 
\psfig{file=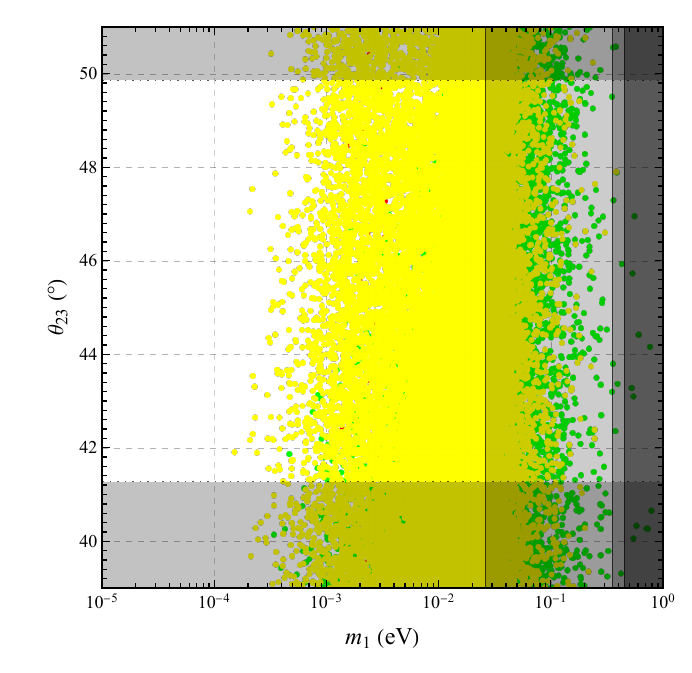,scale=0.2} 
\psfig{file=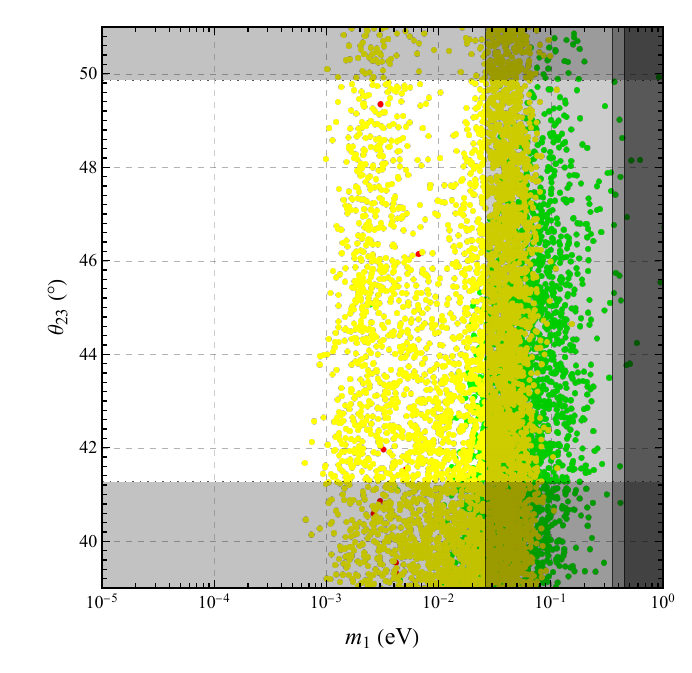,scale=0.2} 
\psfig{file=fA0.pdf,scale=0.2} 
\end{center}
\caption{Scatter plots projected in the plane $\theta_{23}$ versus $m_1$
for an increasing value of $\theta_{23}^{\rm L} \in 10^\circ [n,n+1] (n=0, 17)$,
except for the first and last panel that are for 
the standard case $\theta_{23}^{\rm L} < \theta_{23}^{\rm CKM}$.
We do not show here strong thermal solutions (blue points) for clarity.}
\end{figure}
One can see that starting from the first panel, that is for the standard case when $\theta_{23}^{\rm L} < \theta_{23}^{\rm CKM}$,
by increasing $\theta_{23}^{\rm L}$ one can notice how the muonic solutions become more and more dominant and 
their lower bound on $m_1$ decreases. Already in the fourth panel, for $\theta_{23}^{\rm L} \in [20^\circ,30^\circ]$,
the absolute lower bound on $m_1$ is saturated by muonic solutions. Their dominance grows up to the panel for 
$\theta_{23}^{\rm L} \in [40^\circ,50^\circ]$. After the panel for the interval 
$\theta_{23}^{\rm L} \in [80^\circ,90^\circ]$, the scatter plots are clearly symmetric so that the initial case is restored in the final (20th)
panel. These numerical results can be simply explained by the analytical expression for the flavoured $C\!P$ asymmetries. 

\section{Final remarks}

We have extended the traditional results from $SO(10)$-inspired leptogenesis to
the case when the unitary matrix transforming the LH neutrino from the weak basis to the neutrino
Yukawa basis is left arbitrary. Let us make some final remarks. 
\begin{itemize}
\item We have shown that, given current neutrino experimental data,
just imposing that the neutrino Dirac masses are not too dissimilar from the up quark masses yields
a RH neutrino mass spectrum with $M_1 < 10^9\,{\rm GeV}$ and $M_2 \sim 10^{11}\,{\rm GeV}$,
necessarily requiring $N_2$-leptogenesis for the explanation of the measured baryon asymmetry. 
The only way to avoid this conclusion, barring special crossing level solutions or very particular choices for
the matrix $V_L$, is that either $\alpha_1 \gtrsim 100$ or that $\alpha_1 \lesssim 0.01$. 
In this first case the $N_1$'s become heavier than $~10^{9}\,{\rm GeV}$ and one recovers a $N_1$-leptogenesis scenario 
without any fine-tuning. In the second case it becomes lighter than the sphaleron freeze-out temperature, $\sim 100\,{\rm GeV}$,
and one recovers an effective 2 RH neutrino scenario \cite{Antusch:2011nz}, as noticed in \cite{DiBari:2015svd}.  
\item  We have found that in extended $SO(10)$-inspired leptogenesis 
the solutions fill all the region in the space $m_1-\delta-\theta_{23}$ for 
$10^{-4}\,{\rm eV} \lesssim m_1 \lesssim 0.1\,{\rm eV}$ both for NO and IO.
 However, a lower bound on $m_1$, though relaxed by about two orders of magnitude,
still holds. 
\item  Imposing in addition a strong thermal leptogenesis condition,  IO is still not viable. 
This shows how NO is quite an unavoidable condition for strong thermal $SO(10)$-inspired leptogenesis, 
even within the extended version we have considered here. 
The current support of data to NO should then regarded as an encouraging success for this scenario.
\item  At the same, even without strong thermal condition, an ultimate determination of NO would 
certainly be in favour of a strict $SO(10)$-inspired scenario with no or partially extended $V_L$.
In this respect we have studied in detail how conditions for IO solutions to appear require a big
departure of $V_L$ from the CKM matrix. 
\item Finally, we would like to notice that thought the lower bound on $m_{ee}$
is relaxed by one order of magnitude in this extended version of $SO(10)$-inspired leptogenesis, still
the bulk of solutions ($~ 95\%$) lie in a range between 1 meV and 30 meV. This means 
that in future years new generation $0\nu\b\b$ experiments will be able to test extended $SO(10)$-inspired leptogenesis
in a substantial way. Furthermore,  the strong thermal subset of the solutions would imply that a signal 
should be found by next generation experiments: it is certainly an exciting time for $SO(10)$-inspired leptogenesis. 
\end{itemize}

\vspace{-1mm}
\subsection*{Acknowledgments}

We acknowledge financial support from the STFC Consolidated Grant ST/T000775/1.
PDB also acknowledges support from the European Union’s Horizon 2020 Europe research and innovation programme under  
the Marie Sk\l odowska-Curie grant agreement HIDDeN European  ITN project (H2020-MSCA-ITN2019//860881-HIDDeN).
We wish to thank Franco Buccella, Frank Deppisch, Steve King and Frederick Wehlen for comments and discussions. 


\renewcommand{\thesection}{\Alph{section}}
\renewcommand{\thesubsection}{\Alph{section}.\arabic{subsection}}
\def\theequation{\Alph{section}.\arabic{equation}}
\renewcommand{\thetable}{\arabic{table}}
\renewcommand{\thefigure}{\arabic{figure}}
\setcounter{section}{1}
\setcounter{equation}{0}

\end{document}